\newcommand{\be}{\begin{equation}}
\newcommand{\ee}{\end{equation}}
\newcommand{\bea}{\begin{eqnarray}}
\newcommand{\eea}{\end{eqnarray}}
\newcommand{\beas}{\begin{eqnarray*}}
\newcommand{\eeas}{\end{eqnarray*}}
\newcommand{\nn}{\nonumber}

\documentclass[onecolumn,aps,showpacs,nofootinbib,epsfig,axodraw]{revtex4}
\usepackage{graphics}
\usepackage{epsfig}
\usepackage{axodraw}
\usepackage[dvips]{color}

\def\primero{\begin{picture}(100,90)(-25,-35)
\SetWidth{1.2} \SetScale{0.4}\Text(-12,45)[]{$D_1$}
\Vertex(46,15){2} \Vertex(86,15){2} \SetColor{Red}
\Photon(46,15)(-16,75){4}{9} \Photon(86,15)(146,75){4}{9}
\PhotonArc(66,15)(20,180,360){3}{9} \SetColor{Blue}
\CArc(66,15)(20,0,180) \ArrowLine(-16,-45)(46,15)
\LongArrow(86,15)(146,-45) \Text(65,32)[]{$k'$} \Text(-12,32)[]{$k$}
\Text(-12,-20)[]{$p$} \Text(65,-20)[]{$p'$}
\end{picture}}

\def\segundo{\begin{picture}(100,90)(-25,-35)
\SetWidth{1.2} \SetScale{0.4} \Text(-12,45)[]{$D_2$}
\Vertex(46,15){2} \Vertex(86,15){2} \SetColor{Red}
\Photon(95,95)(42,15){4}{11} \Photon(90.5,15)(52,95){4}{11}
\PhotonArc(66,15)(20,180,360){3}{9} \SetColor{Blue}
\CArc(66,15)(20,0,180) \ArrowLine(6,-45)(46,15)
\LongArrow(86,15)(126,-45) 
\end{picture}}

\def\tercero{\begin{picture}(100,90)(-25,-53)
\SetWidth{1.2} \SetScale{0.4} \Text(-12,25)[]{$D_3$}
\Vertex(66,15){2} \Vertex(66,-55){2} \Vertex(66,-25){2}
\SetColor{Red} \Photon(66,15)(6,75){4}{9}
\Photon(66,15)(126,75){4}{9} \Photon(66,-25)(66,-55){4}{4}
\SetColor{Blue} \CArc(66,-5)(20,0,360) \ArrowLine(6,-125)(66,-55)
\LongArrow(66,-55)(126,-125) 
\end{picture}}

\def\cuarto{\begin{picture}(100,90)(-30,-35)
\SetWidth{1.2} \SetScale{0.4} \Text(-16,43)[]{$D_4$}
\Vertex(46,15){2} \Vertex(86,15){2} \Vertex(16,15){2} \SetColor{Red}
\Photon(16,15)(-46,75){4}{9} \Photon(86,15)(146,75){4}{9}
\PhotonArc(66,15)(20,0,180){3}{9} \SetColor{Blue}
\Line(16,15)(46,15) \CArc(66,15)(20,180,360)
\ArrowLine(-46,-45)(16,15) \LongArrow(86,15)(146,-45)
\end{picture}}

\def\quinto{\begin{picture}(100,90)(-27,-25)
\SetWidth{1.2}  \SetScale{0.4} \Text(-12,55)[]{$D_5$}
\Vertex(46,15){2} \Vertex(86,15){2} \Vertex(16,15){2} \SetColor{Red}
\Photon(16,15)(75,125){4}{14} \Photon(90.5,15)(45,125){4}{14}
\PhotonArc(66,15)(20,0,180){3}{9} \SetColor{Blue}
\Line(16,15)(46,15) \CArc(66,15)(20,180,360)
\ArrowLine(-26,-45)(16,15) \LongArrow(86,15)(126,-45)
\end{picture}}

\def\sexto{\begin{picture}(100,90)(-27,-25)
\SetWidth{1.2} \SetScale{0.4} \Text(-12,55)[]{$D_6$}
\Vertex(56,15){2} \Vertex(86,15){2} \Vertex(16,15){2} \SetColor{Red}
\Photon(13.5,15)(65,125){4}{14} \Photon(90.5,15)(35,125){4}{14}
\PhotonArc(36,15)(20,180,360){3}{9} \SetColor{Blue}
\Line(56,15)(86,15) \CArc(36,15)(20,0,180)
\ArrowLine(-26,-45)(16,15) \LongArrow(86,15)(126,-45)
\end{picture}}

\def\septimo{\begin{picture}(100,90)(-30,-35)
\SetWidth{1.2} \SetScale{0.4} \Text(-15,45)[]{$D_7$}
\Vertex(56,15){2} \Vertex(86,15){2} \Vertex(16,15){2} \SetColor{Red}
\Photon(16,15)(-46,75){4}{9} \Photon(86,15)(146,75){4}{9}
\PhotonArc(36,15)(20,180,360){3}{9} \SetColor{Blue}
\Line(56,15)(86,15) \CArc(36,15)(20,0,180)
\ArrowLine(-46,-45)(16,15) \LongArrow(86,15)(146,-45)
\end{picture}}

\def\octavo{\begin{picture}(100,90)(-30,-37)
\SetWidth{1.2} \SetScale{0.4}  \Text(-15,43)[]{$D_8$}
\Vertex(46,15){2} \Vertex(21,-10){2} \Vertex(71,-10){2}
\SetColor{Red} \Photon(46,15)(-16,75){4}{9}
\Photon(46,15)(111,80){4}{9} \Photon(21,-10)(71,-10){4}{6}
\SetColor{Blue} \ArrowLine(-16,-45)(46,15)
\LongArrow(46,15)(111,-50) 
\end{picture}}

\def\noveno{\begin{picture}(100,90)(-25,-40)
\SetWidth{1.2} \SetScale{0.4} \Text(-12,40)[]{$D_9$}
\Vertex(21,25){2} \Vertex(21,-10){2} \Vertex(81,0){2} \SetColor{Red}
\Photon(21,25)(-6,85){4}{8} \Photon(81,0)(121,85){4}{10}
\Photon(21,-10)(81,0){4}{7} \SetColor{Blue}
\ArrowLine(21,-10)(21,25) \ArrowLine(21,25)(81,0)
\ArrowLine(-6,-75)(21,-10) \LongArrow(81,0)(121,-75)
\end{picture}}

\def\decimo{\begin{picture}(100,90)(-25,-40)
\SetWidth{1.2} \SetScale{0.4} \Text(-8,40)[]{$D_{10}$}
\Vertex(21,25){2} \Vertex(21,-10){2} \Vertex(81,0){2} \SetColor{Red}
\Photon(21,25)(116,85){4}{13} \Photon(81,0)(46,85){4}{10}
\Photon(21,-10)(81,0){4}{7} \SetColor{Blue}
\ArrowLine(21,-10)(21,25) \ArrowLine(21,25)(81,0)
\ArrowLine(-6,-75)(21,-10) \LongArrow(81,0)(121,-75)
\end{picture}}

\def\onceavo{\begin{picture}(100,90)(-53,-40)
\SetWidth{1.2} \SetScale{0.4} \Text(-35,40)[]{$D_{11}$}
\Vertex(21,25){2} \Vertex(21,-10){2} \Vertex(-41,0){2}
\SetColor{Red} \Photon(21,25)(-21,85){4}{8}
\Photon(-41,0)(26,85){4}{13} \Photon(21,-10)(-41,0){4}{7}
\SetColor{Blue} \ArrowLine(21,25)(21,-10) \ArrowLine(-41,0)(21,25)
\LongArrow(21,-10)(56,-75) \ArrowLine(-81,-75)(-41,0)
\end{picture}}

\def\doce{\begin{picture}(100,90)(-53,-40)
\SetWidth{1.2}  \SetScale{0.4} \Text(-35,40)[]{$D_{12}$}
\Vertex(21,25){2} \Vertex(21,-10){2} \Vertex(-41,0){2}
\SetColor{Red} \Photon(21,25)(56,85){4}{8}
\Photon(-41,0)(-81,85){4}{10} \Photon(21,-10)(-41,0){4}{7}
\SetColor{Blue} \ArrowLine(21,25)(21,-10) \ArrowLine(-41,0)(21,25)
\LongArrow(21,-10)(56,-75) \ArrowLine(-81,-75)(-41,0)
\end{picture}}

\def\trece{\begin{picture}(100,90)(-27,-53)
\SetWidth{1.2} \SetScale{0.4} \Text(-12,26)[]{$D_{13}$}
\Vertex(66,15){2} \Vertex(66,-55){2} \Vertex(26,35){2}
\SetColor{Red} \Photon(26,35)(-6,75){4}{4.5}
\Photon(66,15)(126,75){4}{9} \Photon(66,15)(66,-55){4}{8}
\SetColor{Blue} \CArc(46,25)(20,0,360) \ArrowLine(6,-125)(66,-55)
\LongArrow(66,-55)(126,-125) 
\end{picture}}

\def\catorce{\begin{picture}(100,90)(-25,-53)
\SetWidth{1.2} \SetScale{0.4} \Text(-10,26)[]{$D_{14}$}
\Vertex(66,15){2} \Vertex(66,-55){2} \Vertex(96,45){2}
\SetColor{Red} \Photon(66,15)(-6,75){4}{9}
\Photon(96,45)(136,75){4}{4.5} \Photon(66,15)(66,-55){4}{8}
\SetColor{Blue} \CArc(77,35)(20,0,360) \ArrowLine(6,-125)(66,-55)
\LongArrow(66,-55)(126,-125) 
\end{picture}}

\def\quince{\begin{picture}(100,90)(-33,-40)
\SetWidth{1.2} \SetScale{0.4} \Text(-16,38)[]{$D_{15}$}
\Vertex(66,15){2} \Vertex(96,-15){2} \Vertex(16,15){2}
\SetColor{Red} \Photon(16,15)(-46,75){4}{9}
\Photon(66,15)(126,75){4}{9} \PhotonArc(81,0)(20,315,495){3}{9}
\SetColor{Blue} \ArrowLine(16,15)(66,15) \CArc(81,0)(20,495,675)
\ArrowLine(-46,-65)(16,15) \LongArrow(96,-15)(136,-65)
\end{picture}}

\def\diezyseis{\begin{picture}(100,90)(-30,-40)
\SetWidth{1.2} \SetScale{0.4} \Text(-15,38)[]{$D_{16}$}
\Vertex(66,15){2} \Vertex(96,-15){2} \Vertex(16,15){2}
\SetColor{Red} \Photon(16,15)(76,75){4}{9}
\Photon(66,15)(31,75){4}{9} \PhotonArc(81,0)(20,315,495){3}{9}
\SetColor{Blue} \ArrowLine(16,15)(66,15) \CArc(81,0)(20,495,675)
\ArrowLine(-26,-65)(16,15) \LongArrow(96,-15)(116,-65)
\end{picture}}

\def\diezysiete{\begin{picture}(100,90)(-35,-40)
\SetWidth{1.2} \SetScale{0.4} \Text(-18,38)[]{$D_{17}$}
\Vertex(66,15){2} \Vertex(-16,-15){2} \Vertex(15,15){2}
\SetColor{Red} \Photon(16,15)(66,75){4}{9}
\Photon(66,15)(21,75){4}{9} \PhotonArc(-1,0)(20,225,405){3}{9}
\SetColor{Blue} \ArrowLine(16,15)(66,15) \CArc(-1,0)(20,45,225)
\ArrowLine(-46,-65)(-16,-15) \LongArrow(66,15)(116,-65)
\end{picture}}

\def\diezyocho{\begin{picture}(100,90)(-34,-40)
\SetWidth{1.2} \SetScale{0.4} \Text(-18,38)[]{$D_{18}$}
\Vertex(66,15){2} \Vertex(-16,-15){2} \Vertex(15,15){2}
\SetColor{Red} \Photon(16,15)(-46,75){4}{9}
\Photon(66,15)(126,75){4}{9} \PhotonArc(-1,0)(20,225,405){3}{9}
\SetColor{Blue} \ArrowLine(16,15)(66,15) \CArc(-1,0)(20,45,225)
\ArrowLine(-46,-65)(-16,-15) \LongArrow(66,15)(126,-65)
\end{picture}}

\def\diezynueve{\begin{picture}(100,90)(-24,-36)
\SetWidth{1.2} \SetScale{0.4} \Text(-8,41)[]{$D_{19}$}
\Vertex(46,15){2} \Vertex(86,15){2} \Vertex(26,15){2}
\Vertex(106,15){2} \SetColor{Red} \Photon(26,15)(-36,75){4}{9}
\Photon(106,15)(166,75){4}{9} \PhotonArc(66,15)(20,180,360){3}{9}
\SetColor{Blue} \CArc(66,15)(20,0,180) \Line(26,15)(46,15)
\Line(86,15)(106,15) \ArrowLine(-36,-45)(26,15)
\LongArrow(106,15)(166,-45)
\end{picture}}

\def\veinte{\begin{picture}(100,90)(-26,-42)
\SetWidth{1.2} \SetScale{0.4} \Text(-11,36)[]{$D_{20}$}
\Vertex(16,15){2} \Vertex(16,-45){2} \Vertex(16,-15){2}
\Vertex(16,45){2} \SetColor{Red} \PhotonArc(16,0)(15,270,450){3}{7}
\Photon(16,45)(-16,75){4}{4} \Photon(16,-45)(131,75){4}{14}
\SetColor{Blue} \CArc(16,0)(15,90,270) \LongArrow(16,45)(131,-75)
\ArrowLine(-16,-75)(16,-45) \ArrowLine(16,15)(16,45)
\ArrowLine(16,-45)(16,-15) 
\end{picture}}

\def\veinteyuno{\begin{picture}(100,90)(-32,-43)
\SetWidth{1.2} \SetScale{0.4} \Text(-17,36)[]{$D_{21}$}
\Vertex(46,5){2} \Vertex(16,40){2} \Vertex(81,40){2}
\Vertex(46,-35){2} \SetColor{Red} \Photon(16,40)(-26,80){4}{7}
\Photon(81,40)(121,80){4}{7} \Photon(46,5)(46,-35){4}{5}
\SetColor{Blue} \ArrowLine(-16,-95)(46,-35)
\LongArrow(46,-35)(111,-95) \ArrowLine(81,40)(16,40)
\ArrowLine(16,40)(46,5)
\ArrowLine(46,5)(81,40) 
\end{picture}}

\def\veinteydos{\begin{picture}(100,90)(-32,-43)
\SetWidth{1.2} \SetScale{0.4} \Text(-17,36)[]{$D_{22}$}
\Vertex(46,5){2} \Vertex(16,40){2} \Vertex(81,40){2}
\Vertex(46,-35){2} \SetColor{Red} \Photon(16,40)(-26,80){4}{7}
\Photon(81,40)(121,80){4}{7} \Photon(46,5)(46,-35){4}{5}
\SetColor{Blue} \ArrowLine(-16,-95)(46,-35)
\LongArrow(46,-35)(111,-95) \ArrowLine(16,40)(81,40)
\ArrowLine(46,5)(16,40)
\ArrowLine(81,40)(46,5) 
\end{picture}}

\def\veinteytres{\begin{picture}(100,90)(-62,-40)
\SetWidth{1.2} \SetScale{0.4} \Text(-46,39)[]{$D_{23}$}
\Vertex(21,25){2} \Vertex(21,-10){2} \Vertex(-41,0){2}
\Vertex(-71,0){2} \SetColor{Red} \Photon(21,25)(56,85){4}{8}
\Photon(-71,0)(-111,85){4}{10} \Photon(21,-10)(-41,0){4}{7}
\SetColor{Blue} \ArrowLine(-71,0)(-41,0) \ArrowLine(21,25)(21,-10)
\ArrowLine(-41,0)(21,25) \LongArrow(21,-10)(56,-75)
\ArrowLine(-111,-75)(-71,0) 
\end{picture}}

\def\veinteycuatro{\begin{picture}(100,90)(-60,-40)
\SetWidth{1.2} \SetScale{0.4} \Text(-44,38)[]{$D_{24}$}
\Vertex(21,25){2} \Vertex(21,-10){2} \Vertex(-41,0){2}
\Vertex(-71,0){2} \SetColor{Red} \Photon(21,25)(-61,85){4}{8}
\Photon(-71,0)(11,85){4}{10} \Photon(21,-10)(-41,0){4}{7}
\SetColor{Blue} \ArrowLine(-71,0)(-41,0) \ArrowLine(21,25)(21,-10)
\ArrowLine(-41,0)(21,25) \LongArrow(21,-10)(46,-75)
\ArrowLine(-101,-75)(-71,0) 
\end{picture}}

\def\veinteycinco{\begin{picture}(100,90)(-20,-40)
\SetWidth{1.2} \SetScale{0.4} \Text(-5,37)[]{$D_{25}$}
\Vertex(21,25){2} \Vertex(21,-10){2} \Vertex(81,0){2}
\Vertex(111,0){2} \SetColor{Red} \Photon(21,25)(101,85){4}{8}
\Photon(111,0)(11,85){4}{10} \Photon(21,-10)(81,0){4}{7}
\SetColor{Blue} \ArrowLine(81,0)(111,0) \ArrowLine(21,-10)(21,25)
\ArrowLine(21,25)(81,0) \ArrowLine(-6,-75)(21,-10)
\LongArrow(111,0)(151,-75)
\end{picture}}

\def\veinteyseis{\begin{picture}(100,90)(-21,-40)
\SetWidth{1.2} \SetScale{0.4} \Text(-6,38)[]{$D_{26}$}
\Vertex(21,25){2} \Vertex(21,-10){2} \Vertex(81,0){2}
\Vertex(111,0){2} \SetColor{Red} \Photon(21,25)(-6,85){4}{8}
\Photon(111,0)(151,85){4}{10} \Photon(21,-10)(81,0){4}{7}
\SetColor{Blue} \ArrowLine(81,0)(111,0) \ArrowLine(21,-10)(21,25)
\ArrowLine(21,25)(81,0) \ArrowLine(-6,-75)(21,-10)
\LongArrow(111,0)(151,-75)
\end{picture}}

\def\veinteysiete{\begin{picture}(100,90)(55,-107)
\SetWidth{1.2} \SetScale{0.4} \Text(71,-30)[]{$D_{27}$}
\Vertex(226,-185){2} \Vertex(226,-135){2} \Vertex(296,-185){2}
\Vertex(296,-135){2} \SetColor{Red} \Photon(166,-85)(226,-135){4}{7}
\Photon(296,-135)(356,-85){4}{7} \Photon(296,-185)(226,-185){4}{6.5}
\SetColor{Blue} \ArrowLine(166,-235)(226,-185)
\ArrowLine(226,-185)(226,-135) \ArrowLine(226,-135)(296,-135)
\LongArrow(296,-185)(356,-235)  
\ArrowLine(296,-135)(296,-185)
\end{picture}}

\def\veinteyocho{\begin{picture}(100,90)(55,-104)
\SetWidth{1.2} \SetScale{0.4} \Text(71,-28)[]{$D_{28}$}
\Vertex(226,-185){2} \Vertex(226,-135){2} \Vertex(296,-185){2}
\Vertex(296,-135){2} \SetColor{Red} \Photon(296,-65)(226,-135){4}{9}
\Photon(296,-135)(226,-65){4}{9} \Photon(296,-185)(226,-185){4}{6.5}
\SetColor{Blue} \ArrowLine(186,-235)(226,-185)
\ArrowLine(226,-185)(226,-135) \ArrowLine(226,-135)(296,-135)
\LongArrow(296,-185)(336,-235) 
\ArrowLine(296,-135)(296,-185)
\end{picture}}

\begin{document}
\title{On the Compton scattering vertex for massive scalar QED}
\author{A. Bashir$^1$, Y. Concha-S\'anchez$^1$, R. Delbourgo$^2$ and
        M. E. Tejeda-Yeomans$^3$}
\affiliation{$^1$Instituto de F{\'\i}sica y Matem\'aticas,
Universidad Michoacana de San Nicol\'as de Hidalgo, Apartado Postal
2-82, Morelia, Michoac\'an 58040, M\'exico.\\
$^2$School of Mathematics and Physics, University of
Tasmania, Locked Bag 37 GPO, Hobart 7001, Australia. \\
$^3$ Departamento de F{\'\i}sica, Universidad de Sonora, Apartado
Postal 1626, Hermosillo, Sonora 83000, M\'exico.}

\begin{abstract}

We investigate the Compton scattering vertex of charged scalars and photons in
scalar quantum electrodynamics (SQED). We carry out its non perturbative
construction consistent with Ward-Fradkin-Green-Takahashi identity (WFGTI) which
relates 3-point vertices to the 4-point ones.
There is an undetermined part
which is transverse to one or both the external photons, and needs to be
evaluated through perturbation theory. We present in detail how the transverse 
part at  the 1-loop order can be evaluated for completely general 
kinematics of momenta involved in covariant gauges and dimensions. This
involves the calculation of genuine 4-point functions with three massive 
propagators, the most non-trivial integrals reported in this paper.
We also discuss possible applications of our results.

\end{abstract}


\maketitle

\section{Introduction}


Evaluating Green functions
for gauge theories in the non perturbative regime is as important 
as it is
difficult. For example, understanding confinement and dynamical
mass generation in quantum chromodynamics (QCD) requires knowledge 
of quark and gluon propagators in the infra red. These objects are 
tied to the 3- and 4-point
vertices through Schwinger-Dyson equations (SDEs) as well as 
Slavnov-Taylor identities (STI),~\cite{Slavnov:1972-STI}. For example, 
unlike abelian gauge
theories, 4-point ghost-ghost-quark-quark scattering kernel 
arises even at the level where these identities relate 2-
and 3-point vertices in covariant gauges. 4-point quark-quark
scattering kernel for off-shell external legs is another important 
quantity. Its infra red behaviour in the coordinate space would tell
us how the strong potential rises for large distances between two static
quarks. Study of the 4-gluon vertex can improve our insight of the
running coupling in the infra red region,~\cite{Fischer:2008-4g}. 
Understanding
non perturbative behaviour of a 4-point function is a daunting task,
so a detailed study of such a function in a relatively simpler 
scalar quantum electrodynamics~(SQED)
in terms of Ward-Fradkin-Green-Takahashi identity~(WFGTI)~\cite{WFGT} 
can provide an important first step towards
much more involved similar functions in QCD.

 We start by providing the most general basis
in terms of which we can expand out the complete Compton scattering
vertex. We use an appropriate set of basis vectors so that the components
of the 4-point vertex longitudinal to the external photons can be readily
identified. By invoking WFGTI, which relates the 4-point Compton
scattering vertex to the 3-point scalar-photon vertex and
the 2-point scalar propagator, we construct the longitudinal
vertex to all orders in perturbation theory in terms of lower
$n$-point functions. This WGTI-conserving construction will enable us
to go to the next order of approximation in SDE studies. In principle,
we shall now be able to truncate it at the level of the 4-point functions
rather than the 3-point functions.
Note that the WFGTI leaves
the transverse part of the vertex undetermined which
has to be evaluated using the brute force of perturbation theory.

Despite the fact that perturbative evaluation
of $n$-point Green functions in gauge theories is receiving more
attention, analytical results for arbitrary gauge and dimensions
with completely off-shell external legs even for
1-loop 3-point vertices for SQED and 
QCD~\cite{Yajaira, Davydychev:1996, Davydychev:2000, Davydychev:2001}
have been reported not so long ago.
Results for spinor QED can be derived by appropriate replacement of
color factors in the quark-gluon vertex. These
perturbative results provide a natural guide to the possible non
perturbative structures of the fermion-boson vertex, structures
which are vital in the reliable truncations of Schwinger-Dyson
equations at the level of the 3-point functions to study confinement, 
dynamical mass generation and hadron spectrum, see for 
example~\cite{Roberts:2002-vertex, Curtis-Pennington:1990, Bashir:1994}.


A natural next step in this direction is the evaluation of a 4-point
function in arbitrary gauge and dimensions for off-shell external
legs. To our knowledge, no such complete calculation exists for any
gauge theory though a lot of work exists which deals with the
calculation of box-diagrams in particular kinematical regimes related
to the problem at hand. For example, a method to calculate massive
Feynman integrals using Mellin-Barnes technique was developed
in~\cite{Davydychev-0} for propagators and triangles with arbitrary
powers of the denominators. The first on shell infra red divergent
scalar box was calculated in~\cite{Beenakker-1} using a mass
regulator for an internal photon; this result has extensively been
employed in many electroweak and heavy quark calculations since
then. Photon-photon scattering box diagrams were treated in~\cite{Davydychev-3}
using different representations for 1-loop integrals.
2-, 3- and 4-point massive integrals for arbitrary momenta and
dimensions were studied in~\cite{Tarasov-0} with explicit examples
of the integrals relevant to Bhabha scattering. The
4-point master integral in $D$ dimensions using Mellin-Barnes technique
was studied in~\cite{Smirnov}.
General techniques for solving $n$-point scalar massive and massless
1-loop integrals were developed
in~\cite{Davydychev-1,Davydychev-2}. In~\cite{Bhabha-1}, the box diagram 
in arbitrary dimension was calculated for on-shell external legs with 
internal propagators of varying masses in connection with the first order 
radiative corrections to Bhabha scattering in $D$ dimensions.
 1-loop amplitudes for 4-point functions with two external massive quarks
and two external massless partons were studied in detail
in~\cite{Korner}. In connection with 2 $ \rightarrow $ 2 QCD scattering amplitudes, 
scalar 1-loop integrals in different kinematical regimes
have been evaluated in \cite{Beenakker-2,Jeppe,Rodrigo}. More recent
work on the subject can be found in~\cite{Ellis}. Our present article
extends these earlier endeavours to all external 
legs being off-shell with completely different degrees of off-shellness.
Moreover, instead of calculating one particular topology, we evaluate
a complete process in a gauge theory (Compton scattering in SQED)
in an arbitrary covariant gauge, calculating all the diagrams contributing
to the process in general number of dimensions $D$.

SQED is a simple gauge theory as
compared to spinor gauge theories (in the sense that there is no
Dirac matrix structure), but the loop integrals have the same basic
topologies as the ones that appear in spinor QED and QCD.
This makes it an attractive theory. Recently, analytical result for the
six-photon helicity amplitudes has been reported for 
SQED,~\cite{Bernicot:2008}.
Apart from the fact that we develop the machinery needed to
obtain the off-shell 1-loop 4-point function, the study of
processes such as Compton scattering
\be e^-(p) + \gamma(k) \to e^-(p') + \gamma(k')
  \nn \ee
has a value in its own right.
For example, the electric and magnetic polarizabilities of a pion
are related to the Compton scattering amplitude at threshold. Due to
the fact that the internal structure of the pion contributes little
to these polarizabilities, 1-loop results are important to
calculate (see~\cite{RCS} and references therein) because these play
a key role in the calibration of high energy colliders. Moreover, in
terms of physical insight, the analysis of processes with virtual or
real photons attached to an electron or a quark line help us
probe the electron or nucleon substructure (see \cite{Maria Elena}
and references therein for a review in such developments). In
particular, the calculation of 1-loop amplitudes with external
massive quarks are crucial to describe heavy quark hadroproduction
and constant theoretical developments have led to numerical or
semi-analytical procedures that provide results with improved
precision. These advances have allowed for a continuous flow of
different mathematical methods enabling scalar and tensor loop
integral evaluation with massive lines~\cite{Maria Elena}.

In this article, in addition to constructing a WFGTI conserving 
longitudinal Compton scattering vertex, we also present its complete 1-loop 
calculation for off-shell external momenta and for arbitrary gauge and
dimensions for massive SQED. This involves 2-, 3-
and 4-point scalar and tensor integrals up to 2 
indices~\footnote{Tensor integrals involving three Lorentz indices also arise
but the two diagrams involving them cancel each other out because of their 
charge
conjugation symmetry.}. The box
integrals involved have at most three massive propagators. We
proceed by writing down the contributing diagrams and calculating
them in terms of scalar and tensor integrals which we list in the
appendices. Due to exchange symmetries, only eleven of the twenty eight
diagrams are independent. All the calculations have been carried out using
FORM or/and Mathematica 6.0. 

We have organized the article as follows: In section II, 
we start out by proposing a convenient basis to decompose the 
Compton scattering vertex into its longitudinal and transverse 
parts. Moreover, we construct the
longitudinal vertex  non perturbatively. It exhausts 5 of the 10 
basis tensors. In Section II, after setting the notation and 
introducing the Feynman rules, we present all the 1-loop
topologies for the Compton scattering and write them in terms of
tensor and scalar integrals to be evaluated. The reduction of the
tensor integrals in terms of scalar ones and their evaluation has
been done in the appendices. Subtraction of the longitudinal
part earlier constructed yields the undetermined transverse vertex 
to the one loop order. This should provide us with a guide towards its 
possible non perturbative extensions.
In section IV, we present our conclusions. The appendices
have been dedicated to the calculation of the necessary integrals
and presenting the 1-loop results for the full 4-point vertex.

\section{WGTI Conserving Longitudinal Vertex }

The WFGTI which relates 3- and 4-point Green functions is~:
\bea k'^{\mu}\Gamma_{\nu
\mu}(p',k';p,k)&=&\Gamma_{\nu}(p + k,p) - \Gamma_{\nu}(p',p' - k)\;, \nn \\
k^{\mu}\Gamma_{\nu \mu}(p',k';p,k)&=&\Gamma_{\nu}(p',p' + k') -
\Gamma_{\nu}(p - k',p)\;,  \label{LLparts} \eea 
where the vertices
involved are full, as shown in the diagram below~: \SetScale{0.7}
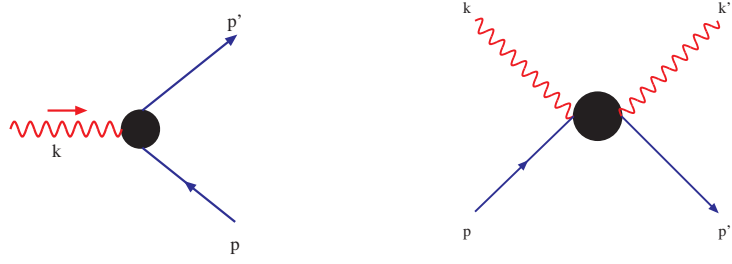
\begin{center}
\begin{picture}(300,70)(0,0)
\SetWidth{1.2} \CCirc(95,15){10}{Black}{Black} \SetColor{Red}
\Photon(25,15)(85,15){4}{7} \LongArrow(45,25)(65,25)
\SetColor{Black} \PText(50,5)(0)[]{k} \PText(146,75)(0)[]{p'}
\PText(146,-45)(0)[]{p}\SetColor{Blue} \LongArrow(96,25)(146,65)
\ArrowLine(146,-35)(96,5)
\end{picture}
\end{center}
\vspace{-1.2cm} \SetScale{0.6}
\begin{center}
\begin{picture}(-100,0)(0,0)
\SetWidth{1.2} \CCirc(66,15){15}{Black}{Black} \SetColor{Red}
\Photon(51,15)(-11,75){4}{9} \Photon(81,15)(141,75){4}{9}
\SetColor{Blue} \LongArrow(81,15)(141,-45)
\ArrowLine(-11,-45)(51,15) \SetColor{Black} \PText(146,85)(0)[]{k'}
\PText(-16,85)(0)[]{k} \PText(-16,-55)(0)[]{p}
\PText(146,-55)(0)[]{p'}
\end{picture}
\end{center}
\vspace{1.5cm} 
The first argument of the 3-point vertex throughout represents
the momentum of the incoming fermion and the second that of the
outgoing one.
The WFGTI identities allow us to write ``longitudinal
parts'' of the 4-point vertex in term of the 3-point vertex. These
parts are defined as the ones which vanish on carrying out any of
the contractions described in~Eq.~(\ref{LLparts}). The remaining
``transverse parts'' will have to be calculated by the brute force
of perturbation theory. We conveniently define 
\bea Q_{\mu}&=&
k_{\mu}'(p + p')\cdot k-k\cdot k'(p + p')_{\mu}  \;,\nn
\\ Q_{\nu}'&=& k_{\nu}(p + p')\cdot k'  - k\cdot k'(p + p')_{\nu}     \;, \nn \\
R_{\mu}&=&k_{\mu}k'\cdot k - k^{2}k_{\mu}'\;, \nn  \\
R_{\nu}'&=&k_{\nu}'k'\cdot k - k'^{2}k_{\nu}\;. \label{BBV} 
\eea
Note that this definition is suitable to separate out the transverse
part of the vertex from the longitudinal one.
Thus the complete 4-point vertex can be written in its 
most general form as~: 
\bea 
\Gamma_{\mu\nu}= \Gamma_{\mu \nu}^L+\Gamma_{\mu
\nu}^T =Ag_{\mu\nu} &+& B_{11}(k\cdot k' g_{\nu\mu} -
k_{\nu}k'_{\mu}) + B_{12}Q_{\nu}'k_{\mu}' +
B_{13}R_{\nu}'k_{\mu}' \nn \\
&+&B_{21}k_{\nu}Q_{\mu} \hspace{1.85cm} + B_{22}Q'_{\nu}Q_{\mu} +
B_{23}R_{\nu}'Q_{\mu} \nn  \\
&+&B_{31}k_{\nu}R_{\mu} \hspace{1.86cm}+ B_{32}Q'_{\nu}R_{\mu} +
B_{33}R_{\nu}'R_{\mu}\;, \label{BBasis} \eea where the coefficients
$A$, $B_{12}$, $B_{21}$, $B_{31}$ and $B_{13}$ are determined by
WFGT identity to all orders in perturbation theory. Therefore, \bea
\Gamma_{\mu \nu}^L &=& Ag_{\mu\nu} + B_{12}Q_{\nu}'k_{\mu}' +
B_{13}R_{\nu}'k_{\mu}' +B_{21}k_{\nu}Q_{\mu}
+B_{31}k_{\nu}R_{\mu}  \;, \\
\Gamma_{\mu \nu}^T &=& B_{11}(k\cdot k' g_{\nu\mu} -
k_{\nu}k'_{\mu})  + B_{22}Q'_{\nu}Q_{\mu} + B_{23}R_{\nu}'Q_{\mu} +
B_{32}Q'_{\nu}R_{\mu} + B_{33}R_{\nu}'R_{\mu}\;. \eea The
coefficients which constitute the longitudinal part of the vertex
are \bea A&=&  - \frac{1}{k\cdot k'} \left\{  \Delta^{-1}(p + k) -
\Delta^{-1}(p) +
\Delta^{-1}(p' - k) - \Delta^{-1}(p') \right\}   \nn \\
B_{21}&=& - \frac{1}{k\cdot k'} \bigg\{\frac{\Delta^{-1}(p + k) -
\Delta^{-1}(p)}{(p + k)^2 -p^2} - \frac{\Delta^{-1}(p') -
\Delta^{-1}(p' - k)}{p'^2 - (p' -
k)^2} - k^2[\Gamma_T(p + k,p) - \Gamma_T(p',p' - k)]\bigg\}  \nn \\
B_{12}&=& - \frac{1}{k\cdot k'}\bigg\{ \frac{\Delta^{-1}(p' - k') -
\Delta^{-1}(p')}{(p' - k')^2 - p'^2} - \frac{\Delta^{-1}(p) -
\Delta^{-1}(p + k')}{p^2 - (p + k')^2} - k'^2
[\Gamma_T(p' - k',p') - \Gamma_T(p,p + k')]\bigg\} \nn \\
B_{31}&=& \frac{1}{(k \cdot k')^2}   \{[(p + k)^2 - p^2]
\Gamma_T(p + k,p) + [(p' - k)^2 -p'^2] \Gamma_T(p',p' - k)\} \nn \\
B_{13}&=&  \frac{1}{(k \cdot k')^2} \{[(p' - k')^2 -p'^2]
\Gamma_T(p' - k',p') + [(p + k')^2 - p^2] \Gamma_T(p,p + k')\} \;,
\label{TLP} \eea where $\Gamma_T$ is the transverse part of the
3-point scalar-photon vertex defined as \be \Gamma_{\mu}(p +k,p)=(2p
+ k)_{\mu}\frac{\Delta^{-1}(p + k) - \Delta^{-1}(p)}{(p + k)^2 -
p^2} + 2(k_{\mu}p\cdot k - k^2p_{\mu}) \Gamma_T(p +k,p)  \label{3PV}
\;. 
\ee 
Note that in all those calculations, where only the
longitudinal projection of the true vertex is
involved,~Eqs.~(\ref{TLP}) constitute the exact non perturbative
result. However, $B_{11}$, $B_{22}$, $B_{23}$, $B_{32}$ and $B_{33}$
remain undetermined and need to be calculated perturbatively order
by order.

In the next section, we calculate the transverse vertex to one loop
order in perturbation theory. This requires the calculation of the full
vertex to that order.

\section{Compton Scattering Vertex}

\subsection{Preliminaries}

   In this sub-section, we shall set the notation and define all the scalar and tensor
integrals to be evaluated. We define the bare quantities in the
usual form:  the scalar propagator $\Delta(p)=  1/(p^2-m^2)$, the
photon propagator $\Delta^0_{\mu \nu} = -\left[ g_{\mu \nu}
p^2-(1-\xi) p_{\mu}p_{\nu}\right]/p^4$, the 3-point vertex
$\Gamma^0_{\mu} = (k+p)_{\mu}$ and the 4-point double photon vertex
$e^2 \Gamma^0_{\mu \nu}= e^2 g_{\mu \nu}$, where $\xi$ is the
general covariant gauge parameter (such that $\xi=0$ corresponds to
Landau gauge) and $e$ is the usual QED coupling constant. Following
is the list of 1-, 2-, 3- and 4-point scalar integrals that we need
to evaluate~: \bea
T &=& \int\frac{d^{D}w}{[ w^{2} - m^{2}]}  \;,   \label{first-scalar}                                   \\
K(k) &=& \int \frac{d^{D}w}{ w^{2}[(k - w)^{2} - m^{2}]} \;,                                          \\
\widetilde{K}(k) &=&  \int \frac{d^{D}w}{[w^{2} - m^{2}][(k - w)^{2} - m^{2}]}  \;,                   \\
L(k) &=& \int \frac{d^{D}w}{ w^{4}[(k - w)^{2} - m^{2}]}  \;,                                         \\
{I}(k,p) &=& \int\frac{d^{D}w}{w^{2} [(k - w)^{2} - m^{2}][(p - w)^{2} - m^{2}]}\;,                 \\
\widetilde{I}(k,p) &=& \int \frac{d^{D}w}{[w^{2}-m^2] [(k - w)^{2} - m^{2}][(p - w)^{2} - m^{2}]}      \\
{J}(k,p) &=& \int \frac{d^{D}w}{w^{4} [(k - w)^{2} - m^{2}][(p - w)^{2} - m^{2}]}  \;,                 \\
{U}(k,p,q) &=& \int \frac{d^{D}w}{w^{2} [(k - w)^{2} - m^{2}][(p - w)^{2} - m^{2}][(q - w)^{2} - m^{2}]} \;, 
\label{Ukpq}  \\
{V}(k,p,q) &=& \int \frac{d^{D}w}{w^{4} [(k - w)^{2} - m^{2}][(p -
w)^{2} - m^{2}][(q - w)^{2} - m^{2}]} \;.
 \label{last-scalar}
\eea
 We also meet tensor integrals up to two indices in such a way that the numerator is $w^{\mu}$ or
$w^{\mu}w^{\nu}$ with the same denominators as in
Eqs.~(\ref{first-scalar}-\ref{last-scalar}). Therefore, we shall
employ the same notation as above for these tensor integrals with
the only difference that we shall add an appropriate tensor
superscript (or  subscript) to the representing symbol.
Thus, for example, \bea
    \widetilde{I}^{\mu \nu}(k,p) &=&\int\frac{w^{\mu} w^{\nu}
 \,\,d^{D}w}{[w^{2}-m^2] [(k - w)^{2} - m^{2}][(p - w)^{2} - m^{2}]}  \;.
\eea
This set of integrals appear in  all the possible 1-loop Feynman
graphs contributing to the Compton scattering process, as we shall see next.
Results of all these integrals are given in detail in the appendices.

Some of these integrals diverge in 4-dimensions and need to be regularized
and renormalized. We do not carry out renormalization in this article but
it is worth mentioning that the scalar integrals $T$, $K(k)$, $L(k)$, $J(k,p)$ and
$V(k,p)$ are divergent, whereas, $I(k,p)$, $\widetilde{I}(k,p)$ and 
$U(k,p,q)$ are convergent. As the tensor integrals can be decomposed in terms of 
scalar integrals, it is easy to deduce their convergence properties.

\subsection{Compton scattering diagrams}


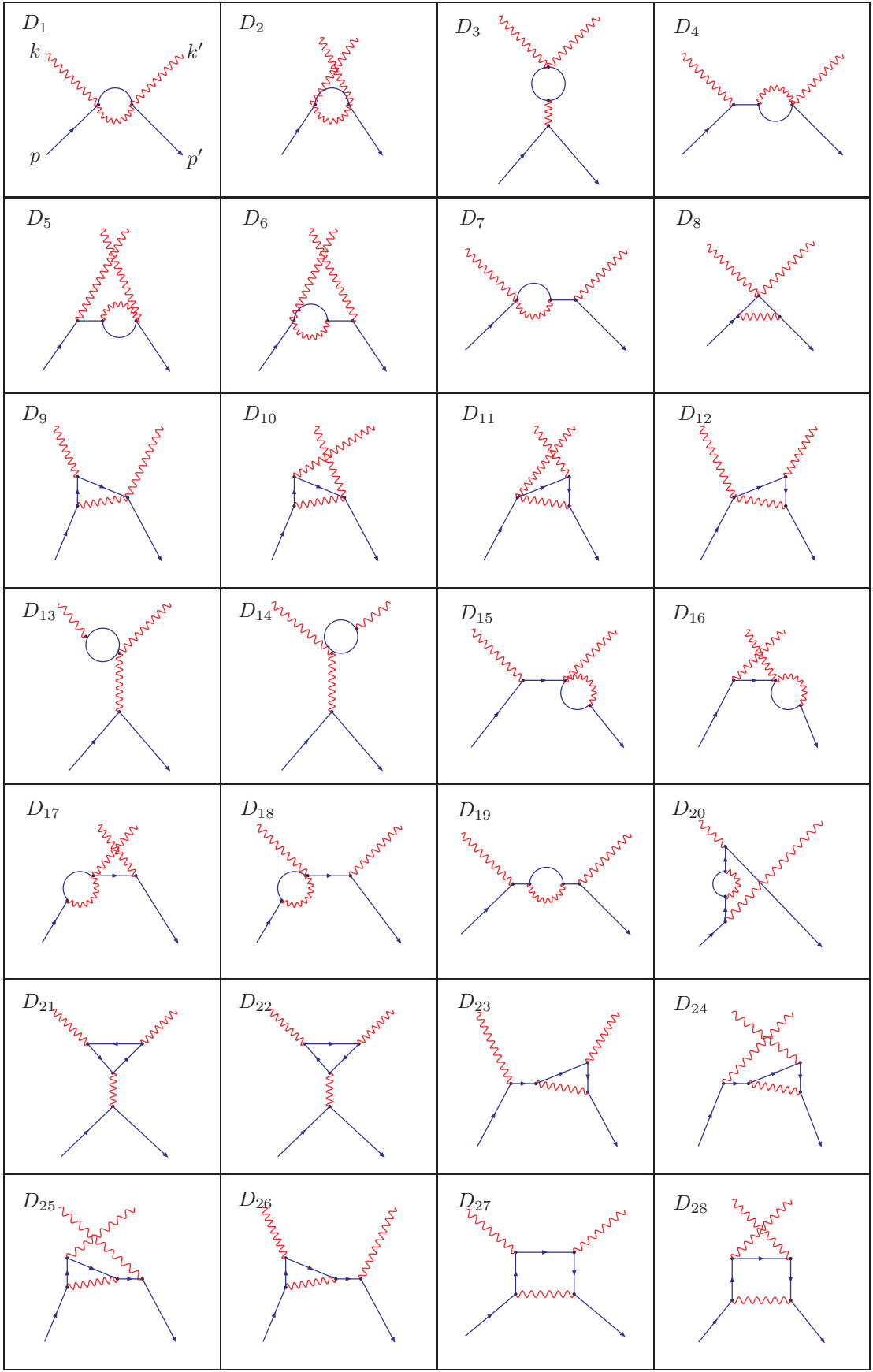
\begin{tabular}{|c|c|c|c|}
\hline
\primero & \segundo & \tercero & \cuarto\\
\hline
\quinto & \sexto & \septimo & \octavo  \\
\hline
\noveno & \decimo & \onceavo & \doce \\
\hline
\trece & \catorce & \quince & \diezyseis \\
\hline
\diezysiete & \diezyocho & \diezynueve & \veinte \\
\hline
\veinteyuno & \veinteydos & \veinteytres & \veinteycuatro  \\
\hline
\veinteycinco & \veinteyseis & \veinteysiete  & \veinteyocho \\
\hline
\end{tabular}

In this
section, we express all these diagrams in terms of the scalar and
multi-indexed integrals defined in the previous section.  Note that
we use the notation  $q=k + p$ and $q'=k' + p$ throughout~: 
\bea \Gamma_{D_1}^{\mu \nu} (q) &=& -\frac{2ie^{2}}{(2\pi)^{D}} \big\{-g^{\mu \nu}K(q) + (1 - \xi)L^{\mu \nu}(q)\big\} \;,
\label{first} \\ \nn \\
     \Gamma_{D_4}^{\mu \nu} (p,q) &=& \frac{ie^{2}}{(2\pi)^{D}}\bigg\{-2\frac{(p^{\mu} +
                               q^{\mu})q^{\nu}}{(q^{2} - m^{2})}K(q) + \frac{(p^{\mu} +
                               q^{\mu})}{(q^{2} - m^{2})}K^{\nu}(q)+ (1 - \xi)(p^{\mu} +
                               q^{\mu})L^{\nu}(q) \bigg\}                                                \;,     \\ \nn \\
     \Gamma_{D_8}^{\mu \nu} &=& \frac{-ie^{2}}{(2\pi)^{D}}g^{\mu\nu}  \bigg\{\widetilde{K}(p - p')
                              - 2(p + p')^{\mu}I_{\mu}(p,p') + 4p\cdot p'I(p,p')  \nn                          \\
                           &&  \hspace{2.0cm} + (\xi - 1)[\widetilde{K}(p-p') - 2(p + p')^{\mu}I_{\mu}(p,p')
                               + 4 p'^{\mu}p^{\nu}J_{\mu \nu}(p,p')] \bigg\}                              \;,     \\ \nn \\
     \Gamma_{D_9}^{\mu \nu} (p,q) &=& \frac{-ie^{2}}{(2\pi)^{D}}\big\{2(p + q)^{\mu}p^{\nu}I(p,q)
                               - (p + q)^{\mu}I^{\nu}(p,q) - 4p^{\nu}I^{\mu}(p,q) + 2I^{\mu \nu}(p,q) \nn       \\
                           &&  \hspace{1.3cm} - (1 - \xi)
                               \big[(p^{2} - m^{2})(p + q)^{\mu}J^{\nu}(p,q) -2(p^{2} -
                               m^{2})J^{\mu \nu}(p,q) \nn       \\
                           &&  \hspace{1.3cm} +2L^{\mu \nu}(q) - (p +q)^{\mu}L^{\nu}(q) \big]\big\}
                            \;,    \\ \nn \\
   \Gamma_{D_{15}}^{\mu \nu} (p,q,p') &=& \frac{ie^{2}}{(2\pi)^{D}}\frac{(p + q)^{\mu}}{(q^{2} - m^{2})}
                               \big\{- 2p'^{\nu}K(p') + K^{\nu}(p') + (1- \xi)(p'^{2} - m^{2})L^{\nu}(p')\big\} \;, \\ \nn \\
   \Gamma_{D_{19}}^{\mu \nu} &=& \frac{ie^{2}}{2(2\pi)^{D}}(p + q)^{\mu}(p' + q)^{\nu}
                               \bigg\{\bigg(\frac{2}{q^{2} - m^{2}} + \frac{4m^{2}}{(q^{2} - m^{2})^{2}}\bigg)
                            K(q)-\frac{T}{(q^{2} - m^{2})^{2}} -\frac{(1 - \xi)}{q^{2} - m^{2}}L(q) \bigg\} \;, \\ \nn \\
   \Gamma_{D_{20}}^{\mu\nu} &=&  \frac{-ie^{2}}{2(2\pi)^{D}}\frac{(p' + q')^{\mu}(p + q')^{\nu}}{(q'^{2}
                              - m^{2})^{2}} \bigg\{ \hspace{-2mm} -4q'^{2}K(q') - T + 4q'_{\alpha}K^{\alpha}(q')
                            + (1 - \xi)\big[ T
                            -4q'_{\alpha}K^{\alpha}(q') \nn \\
                            && \hspace{4.5cm} +4q'_{\alpha}q'_{\beta}
                            L^{\alpha\beta}(q')\big] \bigg\}, \\ \nn \\
  \Gamma_{D_{23}}^{\mu \nu} (p,q,p') &=&\frac{ie^{2}}{2(2\pi)^{D}}\bigg\{(p+q)^{\mu}(p' + q)^{\nu}\bigg[
                             \bigg(\frac{4 p'\cdot q}{q^{2} - m^{2}} + \frac{m^{2} - p'^{2}}{q^{2} - m^{2}} - 1 \bigg)
                             I(p',q)-\frac{\widetilde{K}(p'-q)}{q^{2} - m^{2}}
                              + \frac{K(p')}{q^{2} - m^{2}} \nn \\
                          && + \frac{K(q)}{q^{2} - m^{2}}\bigg] -2(p +
                          q)^{\mu}\bigg[ \bigg(\frac{4 p'\cdot
                            q}{q^{2} - m^{2}} +
                             \frac{m^{2} - p'^{2}}{q^{2} - m^{2}} - 1 \bigg)I^{\nu}(p',q)
                             -\frac{p'^{\nu} + q^{\nu}}{2(q^{2} -
                             m^{2})}\widetilde{K}(p' - q) \nn \\
                          &&   + \frac{K^{\nu}(p')}{q^{2} - m^{2}}+ \frac{K^{\nu}(q)}{q^{2} - m^{2}} \bigg]
                              - (1 - \xi)\bigg[(p + q)^{\mu}(p' +q)^{\nu}\bigg[(p'^{2} - m^{2})J(p',q) - L(q)   \nn \\
                          &&   - \frac{p'^{2} - m^{2}}{q^{2} - m^{2}}L(p')\bigg]
                              -2(p + q)^{\mu}\bigg[(p'^{2} - m^{2})J^{\nu}(p',q) - L^{\nu}(q) -
                              \frac{p'^{2} - m^{2}}{q^{2} - m^{2}}L^{\nu}(p') \bigg] \bigg] \bigg\}    \;, \\ \nn \\
 \Gamma_{D_{27}}^{\mu \nu} (k,p,q,p') &=&\frac{-ie^{2}}{2(2\pi)^{D}}\big\{(k
+ 2p)^{\mu}(k + p + p')^{\nu}[(-2m^{2} + (p' - p)^{2} - 2p\cdot
p')U(p,p',q)+ \widetilde{I}(q
-p,p'-p)  \nn \\
&& - I(p,q) - I(p',q)] - 2(k + p + p')^{\nu}[(-2m^{2} + (p' - p)^{2}
 - 2p\cdot p')U^{\mu}(p,p',q)   \nn \\
&& + \widetilde{I}^{\mu}(q -p,p'-p) - I^{\mu}(p,q) - I^{\mu}(p',q)] - (2k + 4p)^{\mu} [(-2m^{2}+ (p'-
p)^{2} - 2p\cdot p')U^{\nu}(p,p',q) \nn \\ 
&&+ \widetilde{I}^{\nu}(q -p,p'-p) - I^{\nu}(p,q) -I^{\nu}(p',q)] + 4[(-2m^{2} + (p' - p)^{2} -
2p\cdot p')U^{\mu \nu}(p,p',q) \nn \\ 
&&+\widetilde{I}^{\mu \nu}(q -p,p'-p) - I^{\mu \nu}(p,q) -I^{\mu \nu}(p',q)] + (1 -
\xi)\big[(k + 2p)^{\mu}(k + p + p')^{\nu}\big[(-m^{2}(p^{2} + p'^{2}
- m^{2})  \nn \\
&&+ p^{2}p'^{2})V(p,p',q) + (p^{2} + p'^{2} -
2m^{2})U(p,p',q) + \widetilde{I}(q -p,p'-p)  + (m^{2} - p^{2})J(p,q) \nn \\
&&+ (m^{2} - p'^{2})J(p',q) -I(p,q) - I(p',q) +L(q)\big]-2(k +
p + p')^{\nu}\big[(-m^{2}(p^{2} + p'^{2} - m^{2}) \nn \\
&&+ p^{2}p'^{2})V^{\mu}(p,p',q) + (p^{2} + p'^{2} -
2m^{2})U^{\mu}(p,p',q)+ \widetilde{I}^{\mu}(q -p,p'-p)+ (m^{2} -
p^{2})J^{\mu}(p,q) \nn \\
&&+ (m^{2} - p'^{2})J^{\mu}(p',q) -I^{\mu}(p,q) - I^{\mu}(p',q) +L^{\mu}(q) \big] -(2k +
4p)^{\mu}\big[(-m^{2}(p^{2} + p'^{2} - m^{2}) \nn \\
&& + p^{2}p'^{2})V^{\nu}(p,p',q) + (p^{2} + p'^{2} -
2m^{2})U^{\nu}(p,p',q)  + \widetilde{I}^{\nu}(q -p,p'-p) + (m^{2} -
p^{2})J^{\nu}(p,q) \nn \\
&&+ (m^{2} - p'^{2})J^{\nu}(p',q) -I^{\nu}(p,q) - I^{\nu}(p',q) +L^{\nu}(q)\big] + 4\big[(-m^{2}(p^{2} +
p'^{2} - m^{2}) \nn \\
&&+ p^{2}p'^{2})V^{\mu \nu}(p,p',q)  + (p^{2} + p'^{2}
- 2m^{2})U^{\mu \nu}(p,p',q) + \widetilde{I}^{\mu \nu}(q
-p,p'-p) \nn \\
&& + (m^{2} - p^{2})J^{\mu \nu}(p,q) + (m^{2} - p'^{2})J^{\mu
\nu}(p',q) -I^{\mu \nu}(p,q) - I^{\mu \nu}(p',q) + L^{\mu
\nu}(q)\big]\big]\big\}                                         \;.
\label{last}
\eea
The rest of the diagrams can be obtained as follows: \bea
     & \Gamma_{D_2}^{\mu \nu} = \Gamma_{D_1}^{\mu \nu} (q'), &  \\ \nn \\
     \Gamma_{D_5}^{\mu \nu} = \Gamma_{D_4}^{\mu \nu} (p,q'), & 
     \Gamma_{D_6}^{\mu\nu}  = \Gamma_{D_4}^{\mu \nu} (p',q'),&
     \Gamma_{D_7}^{\mu \nu} = \Gamma_{D_4}^{\mu \nu} (p',q), \\ \nn \\
   \Gamma_{D_{10}}^{\mu \nu} = \Gamma_{D_9}^{\mu \nu} (p,q'), & 
   \Gamma_{D_{11}}^{\mu \nu} = \Gamma_{D_9}^{\mu \nu} (p',q'), &
   \Gamma_{D_{12}}^{\mu \nu} = \Gamma_{D_9}^{\mu \nu} (p',q), \\ \nn \\
   \Gamma_{D_{16}}^{\mu \nu} = \Gamma_{D_{15}}^{\mu \nu} (p,q',p'), &
   \Gamma_{D_{17}}^{\mu \nu} = \Gamma_{D_{15}}^{\mu \nu} (p',q',p), &
   \Gamma_{D_{18}}^{\mu \nu} = \Gamma_{D_{15}}^{\mu \nu} (p',q,p), \\ \nn \\
  \Gamma_{D_{24}}^{\mu \nu} = \Gamma_{D_{23}}^{\mu \nu} (p,q',p'), &
  \Gamma_{D_{25}}^{\mu \nu} = \Gamma_{D_{23}}^{\mu \nu} (p',q',p), &
  \Gamma_{D_{26}}^{\mu \nu} = \Gamma_{D_{23}}^{\mu \nu} (p',q,p), \\ \nn \\
  & \Gamma_{D_{28}}^{\mu\nu} = \Gamma_{D_{27}}^{\mu \nu} (k',p,q',p'), &
\eea
while the contributions $\Gamma_{D_3}^{\mu \nu}$, $\Gamma_{D_{13}}^{\mu
  \nu}$, $\Gamma_{D_{14}}^{\mu \nu}$, $\Gamma_{D_{21}}^{\mu\nu} + \Gamma_{D_{22}}^{\mu\nu}$ all vanish.  
Evaluation of the integrals completes the calculation of
the 1-loop Compton scattering amplitude for off-shell external
legs in arbitrary gauge and dimensions in SQED. Results for all the
scalar and multi-indexed tensor integrals have been presented in the
appendices. Through their most general decompositions in terms of
the available vectors, the tensor integrals are converted into a
series of scalar ones for their subsequent evaluation. Most of these
results are new. We verify them to reduce to known expressions for
simpler cases whenever possible. Moreover, as another confirmatory
check, we verify the properties of the scalar pieces which reflect
the symmetries of the original integrals.

 Once we know the full vertex at the one loop order, we can subtract from
it the longitudinal part of the previous section to extract the unknown
transverse coefficients  $B_{11}$, $B_{22}$, $B_{23}$, $B_{32}$ and $B_{33}$
to 1-loop in perturbation theory. Although
straightforward in principle, it is rather involved in practice. The outline 
is as follows. The
4-point Compton scattering vertex can be expanded in a natural basis
as~: \bea \Gamma_{\mu\nu}  \hspace{-0.1cm} = \hspace{-0.1cm} C_0~
g_{\mu\nu} + C_1~ k_{\mu}k_{\nu} + C_2~ k_{\mu}p_{\nu} + C_3~
p_{\mu}k_{\nu} + C_4~ k_{\mu}p'_{\nu} + C_5~ p'_{\mu}k_{\nu}
+ C_6~ p_{\mu}p'_{\nu} + C_7~ p'_{\mu}p_{\nu} + C_8~
p'_{\mu}p'_{\nu} + C_9~ p_{\mu}p_{\nu}, \label{webasis} 
\eea where
the coefficients $C_i$ can be identified from the
expressions~(\ref{first}-\ref{last}). Once the $C_i$ are known, then comparing
Eq.~(\ref{BBasis}) and Eq.~(\ref{webasis}) we can write the unknown
transverse pieces in Eq.~(\ref{BBasis}) at the  1-loop level as
follows~: 
\bea B_{11} &=& - \frac{k^2}{k \cdot k'} ~C_1 + \frac{k^2
(k \cdot p+p^2-p \cdot p')}{(k \cdot k')^2} ~C_2
-\frac{k \cdot p}{k \cdot k'} ~C_3+ \frac{k^2 (-p'^2+p \cdot p'+k \cdot p')}{(k \cdot k')^2} ~C_4 \nn \\
&&-\frac{k \cdot p'}{k \cdot k'} ~C_5 + \frac{k \cdot p (-p'^2+p
\cdot p'+k \cdot p')}{(k \cdot k')^2} ~C_6
+\frac{k \cdot p' (k \cdot p+p^2-p \cdot p')}{(k \cdot k')^2} ~C_7 \nn \\
&&+\frac{k \cdot p' (-p'^2+p \cdot p'+k \cdot p')}{(k \cdot k')^2}
~C_8
+\frac{k \cdot p (k \cdot p+p^2-p \cdot p')}{(k \cdot k')^2} ~C_9 \;, \nn \\
B_{22}  &=& \frac{1}{4}\frac{~C_6+C_7+C_8+C_9}{(k \cdot p')(-2 k
\cdot p +k \cdot p')+(k \cdot p)^2+ k^2 (k^2 +2  k \cdot p
-2 k \cdot p')} \;, \nn \\
B_{23} &=& \frac{1}{4}\frac{C_6-C_7+C_8-C_9}{(k \cdot p')(-2 k \cdot
p+k \cdot p')
+(k \cdot p)^2+k^2(k^2+2 k \cdot p-2 k \cdot p')} \;, \nn \\
B_{32} &=& \frac{1}{4}\frac{2 ~C_2+2 ~C_4-C_6+C_7+C_8-C_9}{(k \cdot
p')(-2 k \cdot p +k \cdot p')+
(k \cdot p)^2+k^2 (k^2+2 k \cdot p-2 k \cdot p')}  \;, \nn \\
B_{33} &=& -\frac{1}{4}\frac{2 ~C_2-2 ~C_4+C_6+C_7-C_8-C_9}{(k \cdot
p')(-2 k \cdot p +k \cdot p')+(k \cdot p)^2+k^2(k^2+2 k \cdot p-2 k
\cdot p')} \;. \label{transverse} \eea 
Thus the decomposition of the
Compton scattering vertex is completely defined by
Eqs.~(\ref{BBV}-\ref{3PV}) and Eq.~(\ref{transverse}). The
longitudinal part has been determined non perturbatively, whereas the
transverse part has been evaluated to 1-loop order. The perturbative
transverse pieces can guide us to their possible non perturbative 
structures which should reduce to this expansion in the weak
coupling regime.

\section{Conclusions}

In this article, we provide a suitable basis to decompose the full Compton 
scattering vertex in SQED into its longitudinal and transverse (to one or
both external photons) parts. We then employ the WFGTI which relates 3-point vertices
to the 4-point ones in order to determine the longitudinal component of the 
Compton scattering vertex in an exact non perturbative fashion, written in 
terms of lower point functions.
Recall that the WGTI-conserving model building for 
the three point interactions provides us with a reliable truncation scheme
for the SDEs at the level of two-point functions. The work presented
in this article provides
an opportunity to lead us to the next order of approximation. One can 
now truncate the tower of SDE at the level of the 4-point function, again with
a WFGTI conserving {\em ansatz} and consistently solve the coupled set of
SDEs for the 2- and the 3-point functions in order to arrive at their
more accurate solutions.
A natural next step is to probe the non perturbative and vitally important 
structures of 4-point
vertices in QCD through the generalized Slavnov-Taylor identities.

As the name suggests, the transverse piece remains undetermined by
the WFGTI. However, with the help of our complete 1-loop evaluation of
this vertex, the unknown transverse pieces can be extracted to that
order. It should serve as a guide for its possible non perturbative
extensions. Any such attempt should reproduce this perturbative result
in the weak coupling regime.

There are further advantages of calculating Compton-scattering for massive
electrons and off-shell external legs in arbitrary gauge and
dimensions:~(i)
As the external legs are off-shell, one can study various cases of
interest by putting the desired particles on shell. On the other
hand, it can also serve as the internal part of more elaborate
Feynman diagrams. (ii) Arbitrary gauge helps in checking the gauge
invariance of the related physical observables and the gauge
covariance properties of the Green functions. (iii) Substituting
$D=4- 2 \epsilon$ and expanding in powers of $\epsilon$, we can
study the 4-dimensional case. Interest has also been shown in lower
dimensional SQED which can be simply projected out by an appropriate
substitution of $D$. (iv) Explicit results of most of the integrals
are new and can also be used in other calculations. These are the
same integrals which are also encountered in spinor QED and QCD. 
Therefore, this study can provide a guideline for a similar
calculation of the ghost-ghost-quark-quark, 4-gluon and 4-quark
vertices in QCD, whose importance has been advocated in the
introductory section.

\section*{ACKNOWLEDGMENTS}

We are highly grateful to A. Davydychev for a very careful reading
of our manuscript and suggesting to us several useful amendments,
corrections and improvements.
We acknowledge CONACyT, COECyT and CIC grants under projects
46614-I and 94527, C8070218-4 and 4.10 respectively.


\section*{APPENDIX I: Master Integrals}

In this appendix we summarize the results for various integrals that
have appeared throughout the calculation. Although some of the
integrals are basic and already known, we tabulate them all for the
sake of completeness. All through the appendix, we  use the
simplifying convention $X_0=(2/i\pi^2) X$ for all scalar integrals
with or without tilde.

\subsection{Scalar Integrals}

\noindent
{\bf The 1- and 2-Point Integrals~:} \\ \\ We start from the list of all the scalar integrals up to 2-point integrals~:
\bea 
T &=& -i\pi^{D/2}(m^{2})^{D/2 -1}\Gamma\bigg(1 - \frac{D}{2}\bigg)  \;,
\\ \nn \\
K(q)&=& -i\pi^{D/2}(m^{2})^{D/2 - 2}
\Gamma\big(1- \frac{D}{2}\big)~
{}_2F_{1}\bigg(2 -\frac{D}{2},1;\frac{D}{2};\frac{q^{2}}{m^{2}}\bigg)  \;,  \\ \nn \\
%
%
%
\widetilde{K}(q) &=&  i \pi^{D/2}\; (m^2)^{D/2-2}\; 
\Gamma\left( 2 - \frac{D}{2}\right)\;
_2F_1\left( 1, 2-\frac{D}{2}; \frac{3}{2}; \frac{q^2}{4m^2} \right)    \;,  \\ \nn \\
L(q)&=&  -i\pi^{D/2}(m^2)^{D/2 -3}
\Gamma \left( 1 - \frac{D}{2}  \right)
{}_2F_{1}\bigg(3 - \frac{D}{2},2
;\frac{D}{2};\frac{q^{2}}{m^{2}}\bigg)  \;. 
\eea
Some of these scalar integrals can be compared with those reported elsewhere.

\begin{itemize}

\item 

One can check that $K(q)$ is the same as $J(\alpha,\beta,0,m)$ in Eq.~(10) of reference~\cite{Davydychev-0}
for $\alpha=\beta=1$.

\item
Similarly $L(q)$ is the same as $J(\alpha,\beta,0,m)$ (as above) for $\alpha=2$ and $\beta=1$. Moreover,
using the recurrence relation (A.17) of~\cite{Smirnov:1996}, one can see
that $L(q)$ is not independent. It can be written in terms of $T$ and $K(q)$ as follows~:
\bea
   L(q) &=& - \frac{1}{(q^2-m^2)^2} \; \left[ (D-3) \left( q^2 + m^2 \right) K(q) - (D-2) T \right] \;.
\eea

\item  Eq.~(17) of~\cite{Davydychev-0} in the case $\nu_1=\nu_2=1$, $p=q$ reproduces 
$\tilde{K}(q)$. It can also be obtained from $J(1,1;m_1,m_2)$, Eqs.~(A.1,~A.7) of~\cite{Smirnov:1996} for
$m_1=m_2$ and $x=y=m^2/q^2$ and after simplification. \\

\end{itemize} 

\noindent
{\bf The 3-Point Integrals~:} \\ \\ We now come to the 3-point function of two variables $k$ and $p$~:
\bea
I(k,p)&=& (m^{2})^{D/2 -3} 
\frac{\Gamma \big(3-\frac{D}{2}\big)}{1-\frac{D}{2}}
\Phi_{2}\Bigg[
\begin{array}{c} 3-\frac{D}{2},1,1,1;\frac{D}{2}-1
\\\frac{D}{2};2 \end{array}\bigg|\frac{k^{2}}{m^{2}},\frac{p^{2}}{m^{2}},\frac{(k-p)^{2}}{m^{2}} \Bigg]  \;,
\\ \nn \\
\widetilde{I}(k,p)&=& - (m^{2})^{D/2 -
3}\frac{\Gamma\big(3 -
\frac{D}{2}\big)}{2}\Phi_{3}\Bigg[\begin{array}{c} 3 -
\frac{D}{2},1,1,1 \\ 3  \end{array} \Bigg| \frac{k^{2}}{m^{2}},\frac{(k-p)^{2}}{m^{2}},\frac{p^{2}}{m^{2}} \Bigg]  \;,
\\ \nn \\
J(k,p)
&\!\!\!=\!\!\!&
\frac{1}{2\chi}
\Biggl\{ 
-2(D-4)(k\!-\!p)^2 \left[m^2+(k p)\right] I(k,p)
+2 (D-3) (k\!-\!p)^2 {\tilde{K}}(k\!-\!p) 
- (D-2) \frac{(k\!-\!p)^2}{m^2} T
\nonumber \\ &&
- \frac{(p^2-m^2)(k-p)^2-2(p^2-k^2)m^2}{(p^2-m^2)^2}
\left[
2(D-3) p^2 K(p) - (D-2)\frac{p^2+m^2}{2m^2} T
\right]
\nonumber \\ &&
- \frac{(k^2-m^2)(k-p)^2+2(p^2-k^2)m^2}{(k^2-m^2)^2}
\left[
2(D-3) k^2 K(k) - (D-2)\frac{k^2+m^2}{2m^2} T
\right]
\Biggr\}  \;,
%
\eea
where 
$\chi=m^2(k^2-p^2)^2 + (m^2-k^2)(m^2-p^2)(k -p)^2$. Note that
the the integrals $I(k,p)$ and $J(k,p)$ are $J_2(1,1,1)$ and $J_2(1,1,2)$ of~\cite{Davydychev:2000}.
Generalized Lauricella function $\Phi_2$ arises for general powers 
of three propagtors $\nu_1$, $\nu_2$ and $\nu_3$ (with the propagator corresponding to $\nu_1$ being
massless, see~\cite{Davydychev-0, Davydychev-1, Davydychev-2} and references therein, for definition, properties and symmetry relations)~:
\bea
\Phi_2\Bigg[\begin{array}{c} \nu_{123} - \frac{D}{2}, \nu_1, \nu_2, \nu_3 ; \frac{D}{2}-\nu_1 \\
\frac{D}{2} ; \nu_{23}  \end{array} \Bigg| x_1,x_2,x_3 \Bigg] &=& 
\sum_{n,l,j=0}^{\infty}  \frac{ x_1^n \;  x_2^l \; x_3^j  }{n! \; l! \; j!} \times    \nn \\ \nn \\ \; 
&& \hspace{-1cm}  \frac{\left( \nu_{123}-\frac{D}{2};n+l+j \right) \left( \nu_1;n+l \right) \left( \nu_2;n+j \right) 
\left( \nu_3;l+j \right) \left( \frac{D}{2} - \nu_1 ; j \right) }{ \left( \frac{D}{2};n+l+j  \right) 
\left( \nu_{23};n+l+2 j \right)  } \;, 
\eea
where the Pochhammer symbol $(z;n)=\Gamma(z+n)/\Gamma(z)$ and $\nu_{i_1,i_2, \cdots, i_j}=\nu_{i_1}+\nu_{i_2}
+ \cdots + \nu_{i_j}$.  Similarly, the generalized Lauricella function 
$\Phi_3$ \footnote{Note that the generalized Lauricella function was first
introduced in~\cite{Srivastava:1969}. Some special cases, including those denoted
as $\Phi_2$ and $\Phi_3$, were employed in~\cite{Davydychev-0}. Further examples
can be found in~\cite{Davydychev-1, Davydychev-2}.} arises for general powers 
of three massive propagtors: $\nu_1$, $\nu_2$ and $\nu_3$,
\bea
\Phi_3\Bigg[\begin{array}{c} \nu_{123} - \frac{D}{2}, \nu_1, \nu_2, \nu_3  \\
 \nu_{123}  \end{array} \Bigg| x_1,x_2,x_3 \Bigg] &=& 
\sum_{n,l,j=0}^{\infty}  \frac{ x_1^n \;  x_2^l \; x_3^j  }{n! \; l! \; j!} 
  \frac{ \left( \nu_{123}-\frac{D}{2};n+l+j \right)  
\left( \nu_1;l+j \right) \left( \nu_2;n+j \right) \left( \nu_3;n+l \right) 
 }{  \left( \nu_{123};n+l+2 j \right)  } \;.
\eea
We finally go on to the 4-point integrals. \\

\noindent
{\bf The 4-Point Integrals~:} \\ \\
\noindent
Genuine 4-point integrals are given by Eq.~(\ref{Ukpq}) and
Eq.~(\ref{last-scalar}). We evaluate them using the Mellin-Barnes technique 
developed in~\cite{Davydychev-0, Davydychev-1, Davydychev-2}\footnote{Reference~\cite{Davydychev-0} develops the 
method. Based upon it, the 4-point functions were considered (along with the arbitrary
number of legs) in~\cite{Davydychev-1} for equal masses and in~\cite{Davydychev-2} for
arbitrary masses.}~:
\bea
U(p,q,p') &=& i \pi^{D/2} (m^{2})^{D/2 -
4}\frac{\Gamma\big(\frac{D}{2} - 1 \big)\Gamma\big(4 -
\frac{D}{2}\big)}{2\Gamma\big(\frac{D}{2}\big)} \times \nn \\ &&
{\cal Y}_{3}   \Bigg[\begin{array}{c} 4 - \frac{D}{2},1,1,1,1,\frac{D}{2} -1 \\
\frac{D}{2} ,3  \end{array} \Bigg| \frac{p^{2}}{m^{2}},
\frac{q^{2}}{m^{2}}, \frac{p'^{2}}{m^{2}},\frac{(q -
p)^{2}}{m^{2}},\frac{(q - p')^{2}}{m^{2}},\frac{(p - p')^{2}}{m^{2}}
\Bigg] \;,   \label{4-point-1}  \\
V(p,q,p') &=&  -i \pi^{D/2}(m^{2})^{D/2 -
5}\frac{\Gamma\big(\frac{D}{2} - 2 \big)\Gamma\big(5 -
\frac{D}{2}\big)}{6\Gamma\big(\frac{D}{2}\big)} \times \nn \\ &&
{\cal Y}_{3}\Bigg[\begin{array}{c} 5 - \frac{D}{2},1,1,1,2,\frac{D}{2} -2 \\
\frac{D}{2} ,4  \end{array} \Bigg| \frac{p^{2}}{m^{2}},
\frac{q^{2}}{m^{2}}, \frac{p'^{2}}{m^{2}},\frac{(q -
p)^{2}}{m^{2}},\frac{(q - p')^{2}}{m^{2}},\frac{(p - p')^{2}}{m^{2}}
\Bigg] \;.   \label{4-point-2} 
\eea 
The scalar integrals $U(p,q,p')$ and $V(p,q,p')$, given by Eqs.~(\ref{4-point-1},~\ref{4-point-2}) 
are genuine 4-point functions with three massive propagators, the most non-trivial integrals 
calculated in this paper. The function ${\cal Y}_3$ is given by
\bea
 {\cal Y}_3\Bigg[\begin{array}{c} \nu_{1234} - \frac{D}{2}, \nu_1, \nu_2, \nu_3, \nu_4 ; \frac{D}{2} - \nu_4  \\
 \frac{D}{2}, \nu_{123}  \end{array} \Bigg| x_1,x_2,x_3,x_4,x_5,x_6 \Bigg] &=& 
\sum_{n,l,r,w,k,j=0}^{\infty}  
\frac{ x_1^n \;  x_2^l \; x_3^r  \; x_4^w \; x_5^k \; x_6^j }{n! \; l! \; r! \; w! \; k! \; j! }  \times \nn \\  
&& \hspace{-10.2cm}
\frac{ \left( \nu_{1234}-\frac{D}{2};n+l+r+w+k+j \right)  
\left( \nu_1;r+k+j \right) 
\left( \nu_2;n+l+r \right) 
\left( \nu_3;n+w+j \right)
\left( \nu_4;l+w+k \right)
\left( \frac{D}{2}- \nu_4;n+r+j \right)
 }{  \left( \frac{D}{2};n+l+r+w+k+j  \right) 
\left( \nu_{123};2n+l+2r+w+k+2j  \right) 
 }.
\eea
This completes the calculation of the scalar integrals.

\subsection{Vector Integrals}

\noindent
The more straightforward vector integrals which are functions of
just one momentum variable are~: 
\bea
 K^{\mu}(q)&=& \frac{q^{\mu}}{2 q^2} \; \left[ T - (m^2-q^2) K(q) \right]  \;, \\ \nn \\
\widetilde{K}^{\mu}(q) &=& \frac{q^{\mu}}{2}  \widetilde{K}(q) \;,  \\ \nn \\
%
%
%
%
L^{\mu}(q) &=& \frac{q^{\mu}}{2q^2}\; \left[ (q^2-m^2)\; L(q)
+ K(q) \right] \;.
%
\eea 
Relatively more involved vector integrals are
functions of two momentum variables which we take up now. \\

\noindent
{\bf The $I^{\mu}(k,p)$ Integral~:} \\

\noindent The  most general description of $I^{\mu}(k,p)$ is~: 
\bea
I_{\mu}(k,p)&=&\frac{i\pi^{2}}{2}[k_{\mu}I_{A}(k,p) +
p_{\mu}I_{B}(k,p)]   \;,
\eea 
where 
\bea \nn
I_{A}(k,p)&=&-\frac{1}{2\Delta^{2}}\bigg\{[p^{2} - (k\cdot
p)]K_{0}(p -k) + [p^{2}(k^{2} - m^2) - k \cdot p(p^2 - m^2)]I_{0}(k,p)
+ p^{2}K_{0}(p) -(k\cdot p)K_{0}(k)\bigg\}  \;,    \\
I_{B}(k,p)&=&I_{A}(p,k)\;,   \label{Itil}  
\eea
and $\Delta^{2}=(k\cdot p)^{2} - k^{2}p^{2}$. \\

\noindent
{\bf The $\widetilde{I}^{\mu}(k,p)$ Integral~:} \\

\noindent Similarly, we can write $\widetilde{I}^{\mu}(k,p)$ as~:
\bea
\widetilde{I}_{\mu}(k,p)&=&\frac{i\pi^{2}}{2}[k_{\mu}\widetilde{I}_{A}(k,p)
+ p_{\mu}\widetilde{I}_{B}(k,p)]   \;,
\eea 
where $\tilde{I}_{A}(k,p)$ and  $\tilde{I}_{B}(k,p)$ are given by the same
expressions as ${I}_{A}(k,p)$ and  ${I}_{B}(k,p)$
in Eqs.~(\ref{Itil}) with the $\; \widetilde{} \;$ now appearing
for each quantity, i.e., with the replacements 
$K_{0}(p -k)  \rightarrow \tilde{K}_{0}(p -k)$,
$I_{0}(k,p) \rightarrow \tilde{I}_{0}(k,p)$, 
$K_{0}(p) \rightarrow  \tilde{K}_{0}(p)  $ 
and 
$K_{0}(k) \rightarrow  \tilde{K}_{0}(k)  $.  \\


\noindent
{\bf The $J^{\mu}(k,p)$ Integral~:} \\

\noindent In the same way, the  most general description of 
$J^{\mu}(k,p)$ is~: \bea
J_{\mu}(k,p)&=&\frac{i\pi^{2}}{2}[k_{\mu}J_{A}(k,p) +
p_{\mu}J_{B}(k,p)]   \;,
\eea 
where $J_{A}(k,p)$ and  $J_{B}(k,p)$ are given by the same
expressions as ${I}_{A}(k,p)$ and  ${I}_{B}(k,p)$ respectively
in Eqs.~(\ref{Itil}) with the replacements 
$  K_{0}(p -k)
\rightarrow I_{0}(k,p)$, $I_{0}(k,p) \rightarrow J_{0}(k,p)$, 
$K_{0}(p) \rightarrow
L_{0}(p)$ and $K_{0}(k) \rightarrow L_{0}(k)$.

We now treat the most complicated vector integrals which depend upon
three
independent momentum variables. \\


\noindent
{\bf The $U^{\mu}(p,q,p')$ Integral~:} \\

\noindent The $U^{\mu}(p,q,p')$ integral can be expressed only in
terms of $p_{\mu}$, $q_{\mu}$ and $p'_{\mu}$. Therefore, \bea
U_{\mu}(p,q,p') &=& \frac{i\pi^{2}}{2}[p_{\mu}U_{A}(p,q,p') +
q_{\mu}U_{B}(p,q,p') + p'_{\mu}U_{C}(p,q,p')] \;,
\eea
where
\bea
U_{A}(p,q,p')&=&\frac{1}{2[p'^{2}(p\cdot q)^{2} - 2(p\cdot q)(p
\cdot p')(p' \cdot q) + q^{2}(p \cdot p')^{2} + p^{2}((p' \cdot
q)^{2} - p'^{2}q^{2})]}\times \nn \\ && \{[(p\cdot p')(p' \cdot q) -
p'^{2}(p \cdot q)]I_{0}(p,p') + [p'^{2}q^{2} - (p'\cdot
q)^{2}]I_{0}(p',q) \nn \\ && + [(p\cdot q)(p' \cdot q)- q^{2}(p\cdot
p')]I_{0}(p,q)+ [p'^{2}(p \cdot q)- p'^{2}q^{2} + q^{2}(p\cdot p')
\nn \\ && - (p\cdot q + p\cdot p')p'\cdot q + (p' \cdot q)^{2}
]\widetilde{I}_{0}(q-p,p'-p) + [p'^{2}q^{2}(-p^{2} + p\cdot q + p\cdot p') \nn
\\ && - p'\cdot q(p'^{2}(p \cdot q)+ q^{2}(p \cdot p')) + p^{2}(p'
\cdot q)^{2} + m^{2}(-p'^{2}(p \cdot q) + p'^{2}q^{2} \nn \\ && -
q^{2}(p \cdot p') + p' \cdot q(p \cdot q + p\cdot p') - (p' \cdot
q)^{2})]U_{0}(p,q,p') \} \;, \nn \\ \nn \\  U_{B}(q,p,p') &=&
U_{C}(p',q,p) = U_{A}(p,q,p')\;.   \label{UiS} \eea

\noindent
{\bf The $V^{\mu}(p,q,p')$ Integral~:} \\

\noindent

\noindent
The integral $V^{\mu}(p,q,p')$ can also be expanded out as follows~:

\bea V_{\mu}=\frac{i\pi^{2}}{2}[p_{\mu}V_{A}(p,q,p') +
q_{\mu}V_{B}(p,q,p') + p'_{\mu}V_{C}(p,q,p')] \;, \eea where
$V_{A}(p,q,p')=V_{B}(q,p,p')=V_{C}(p',q,p)$ are the same as
$U_{A}(p,q,p')=U_{B}(q,p,p')=U_{C}(p',q,p)$ in Eqs.~(\ref{UiS})
with the following replacements~: $I_{0}(p,p')
\rightarrow J_{0}(p,p')$, $I_{0}(p',q) \rightarrow J_{0}(p',q)$,
$I_{0}(p,q) \rightarrow J_{0}(p,q)$, 
$\tilde{I}_{0}(q-p,p'-p) \rightarrow
U_{0}(p,q,p')$ and $U_{0}(p,q,p') \rightarrow V_{0}(p,q,p')$.

\subsection{Tensor Integrals with Two Indices}

\noindent
These integrals can also be categorized as regards the
number of independent momentum variables they depend upon. For only
one momentum dependence, we have
\bea \widetilde{K}^{\mu \nu}(q)&=& \frac{1}{4(D-1)}  
\left[  2(D - 2) T +
\left( q^2 D - 4 m^2 \right)
\widetilde{K}(q)  \right]
\bigg(\frac{q^{\mu}q^{\nu}}{q^2} -
\frac{1}{D}g^{\mu \nu}\bigg) + \frac{g^{\mu \nu}}{D} \left( T+
{m^2} \widetilde{K}(q)\right)\;,  \\ \nn \\
L^{\mu\nu}(q) &=& \frac{1}{4(D-1)q^2}
\left[
(q^2-m^2)^2L(q)-2(q^2+m^2)K(q)+T
\right]
\left(D\frac{q^{\mu}q^{\nu}}{q^2}-g^{\mu\nu}\right)
+ \frac{q^{\mu}q^{\nu}}{q^2} K(q)  \;.
\eea 
Solutions for the integrals with growing
complexity have been listed below
separately~: \\


\noindent
{\bf The $I^{\mu \nu}(k,p)$ Integral~:} \\

\noindent 
The tensor integral $I^{\mu\nu}$ can be expressed in terms
of the scalar integrals $K_{0},I_{C}, I_{D}$ and $I_{E}$ as follows~:
\bea I_{\mu\nu}^{(2)} &=&
\frac{i\pi^{2}}{2}\bigg[\frac{g_{\mu\nu}}{D}K_0(p - k) +
\bigg(k_{\mu}k_{\nu} - g_{\mu\nu}\frac{k^{2}}{D}\bigg)I_C +
\bigg(p_\mu k_\nu + k_\mu p_\nu - g_{\mu\nu}\frac{2(k \cdot
p)}{D}\bigg)I_{D} + \bigg(p_\mu p_\nu
-g_{\mu\nu}\frac{p^{2}}{D}\bigg)I_{E}\bigg]\;. \nn \eea The
coefficients $I_C,\,I_D$ and $I_E$ in the above expressions are~:
\bea I_{C}(k,p)&=&\frac{D}{(D -
2)\Delta^{2}}\bigg\{\bigg[\bigg(\frac{1}{2} -\frac{1}{D}\bigg)(p^{2}
- m^2)k\cdot p - \bigg(\frac{1}{2} -\frac{1}{2D}\bigg)(k^{2} -
m^2)p^{2}\bigg]I_{A} - \frac{1}{2D}(p^{2} - m^2)p^2I_{B}\nn
\\&& \hspace{2cm}
 + \bigg[\bigg(\frac{2}{D} -
\frac{1}{2}\bigg)p^{2} + \bigg(\frac{1}{2} -
\frac{1}{D}\bigg)(k\cdot p)\bigg] \frac{K_{0}(p - k)}{2} -\bigg(1 -
\frac{2}{D}\bigg)p_{\nu}K^{\nu}_{0}(k)\bigg\}\;, \\ \nn \\
I_{D}(k,p)&=&\frac{D}{2(D -
2)\Delta^{2}}\bigg\{\bigg[(k^{2} - m^2)k\cdot p -
\bigg(1 - \frac{2}{D} \bigg)(p^{2} - m^2)k^{2}\bigg] \frac{I_{A}}{2}
+\bigg[(p^{2} - m^2)k\cdot p \nn \\ 
&& \hspace{2cm}-
\bigg(1 - \frac{2}{D} \bigg)(k^{2} - m^2)p^{2}\bigg] \frac{I_{B}}{2} -
\frac{1}{4}\bigg[\bigg(1 - \frac{2}{D}\bigg)(p - k)^{2} +
\frac{4}{D}(k\cdot p)\bigg]K_{0}(p - k)
 \nn \\
&&  \hspace{2cm} +\bigg(1 -
\frac{2}{D}\bigg)
\left[k_{\nu}K^{\nu}_{0}(k) + p_{\nu}K^{\nu}_{0}(p)\right] \bigg\}\;. 
\eea 
Note that the momentum dependence of $I_A$ and $I_B$ has not been displayed
explicitly. It is $I_A(k,p)$ and $I_B(k,p)$. Moreover, $I_{C}(k,p)$
and $I_{E}(k,p)$ are symmetric functions under the exchange of $k$
and $p$. Therefore, \bea I_{E}(k,p)=I_{C}(p,k)\;. \eea


\noindent
{\bf The $J^{\mu \nu}(k,p)$ Integral~:} \\

\noindent The tensor integral $J^{\mu\nu}$ has the following decomposition in terms
of the scalar integrals $I_{0}, J_{C}, J_{D}$ and $J_{E}$~: 
\bea J_{\mu\nu}^{(2)} &=&
\frac{i\pi^{2}}{2}\bigg[\frac{g_{\mu\nu}}{D}I_0 +
\bigg(k_{\mu}k_{\nu} - g_{\mu\nu}\frac{k^{2}}{D}\bigg)J_C +
\bigg(p_\mu k_\nu + k_\mu p_\nu - g_{\mu\nu}\frac{2(k \cdot
p)}{D}\bigg)J_{D} + \bigg(p_\mu p_\nu
-g_{\mu\nu}\frac{p^{2}}{D}\bigg)J_{E}\bigg]\;. \nn \eea The
coefficients $J_C,\,J_D$ and $J_E$ in the above expressions are~:
\bea J_{C}(k,p)&=&\frac{D}{(D - 2)\Delta^{2}}
\bigg\{\bigg[\bigg( \frac{1}{2}
-\frac{1}{D}\bigg)(p^{2} - m^2)k\cdot p - \bigg( \frac{1}{2}
-\frac{1}{2D}\bigg)(k^{2} - m^2)p^{2}\bigg]{J_{A}} -
\frac{1}{2D}(p^{2} - m^2)p^2J_{B}\nn
\\&& \hspace{2cm}
+ \bigg[\bigg(1 - \frac{2}{D}\bigg)k\cdot p - \bigg(1 -
\frac{1}{D}\bigg)p^{2}\bigg] \frac{I_{A}}{2} -\frac{p^2}{2D}I_{B} +
\frac{p^2}{D}I_{0} -\bigg(1 -
\frac{2}{D}\bigg)p_{\nu}L^{\nu}_{0}(k)\bigg\}\;, \\ \nn \\
J_{D}(k,p)&=&\frac{D}{2(D - 2)\Delta^{2}}\bigg\{\bigg[(k^{2} -
m^2)k\cdot p - \bigg(1 - \frac{2}{D}\bigg)(p^{2} -
m^2)k^{2}\bigg]\frac{J_{A}}{2} +\bigg[(p^{2} - m^2)k\cdot p \nn \\
&& \hspace{2cm}- \bigg(1 - \frac{2}{D} \bigg)(k^{2} -
m^2)p^{2}\bigg]\frac{J_{B}}{2} + \bigg[k\cdot p - \bigg(1 -
\frac{2}{D}\bigg)k^{2}\bigg]\frac{I_{A}}{2} + \bigg[k\cdot p -
\bigg(1 - \frac{2}{D}\bigg)p^{2}\bigg]\frac{I_{B}}{2}
 \nn \\
&&  \hspace{2cm} - \frac{2}{D}(k \cdot p)I_{0}+\bigg(1 -
\frac{2}{D}\bigg)\left[k_{\nu}L^{\nu}_{0}(k) +
p_{\nu}L^{\nu}_{0}(p)\right] \bigg\}\;. 
\eea 
The momentum dependence of $I_A, I_B, J_A, J_B$ is 
$I_A(k,p), I_B(k,p), J_A(k,p), J_B(k,p)$.
Moreover, as is evident, $J_{C}(k,p)$ and $J_{E}(k,p)$ are symmetric 
under the exchange of $k$ and $p$, {\it i.e.}, \bea J_{E}(k,p)=J_{C}(p,k)\;. \eea


\noindent
{\bf The $U^{\mu \nu}(p,q,p')$ Integral~:}  \\

\noindent The general expansion of $U^{\mu\nu}$  in terms of the
scalar integrals $\widetilde{I}_{0}, U_{D}, U_{E}$ ,$U_{F}$,
$U_{G}$, $U_{H}$ and $U_{I}$ is as follows~:
\bea
U_{\mu\nu}(p,q,p')&=&\frac{i\pi^{2}}{2}\bigg[\frac{g_{\mu\nu}}{D}
\widetilde{I}_{0}(q-p,p'-p)
+ \bigg(p_{\mu}p_{\nu} - g_{\mu\nu}\frac{p^{2}}{D}\bigg)U_{D} +
\bigg(p_{\mu}q_{\nu} + q_{\mu}p_{\nu} -
g_{\mu\nu}\frac{2(p \cdot q)}{D}\bigg)U_{E} \nn\\
&& + \bigg(p_{\mu}p'_{\nu} + p'_{\mu}p_{\nu} - g_{\mu\nu}\frac{2(p
\cdot p')}{D}\bigg)U_{F} +\bigg(q_{\mu}q_{\nu} -
g_{\mu\nu}\frac{q^{2}}{D}\bigg)U_{G} \nn \\ && +
\bigg(q_{\mu}p'_{\nu} + p'_{\mu}q_{\nu} - g_{\mu\nu}\frac{2(p' \cdot
q)}{D}\bigg)U_{H} + +\bigg(p'_{\mu}p'_{\nu} -
g_{\mu\nu}\frac{p'^{2}}{D}\bigg)U_{I} \bigg]\;,
\eea
where
\bea U_{D}(p,q,p')&=&\frac{1}{2 (D-3)
\left[p^2 (p' \cdot q)^2 - 2 (p \cdot p')( p \cdot q)( p' \cdot
q)+(p \cdot p')^2 q^2+p'^2 \left((p \cdot q)^2-p^2
q^2\right)\right]}\nn \\ && \big \{\big[ (D-2) p^2 (p' \cdot
q)^2-(D-3) (p'^2 p \cdot q+p \cdot p' q^2) p' \cdot q+p'^2 ((D-3) (p
\cdot p'+p \cdot q) \nn \\ && -(D-2) p^2) q^2+m^2 (-(D-2) (p' \cdot
q)^2+(D-3) (p \cdot p'+p \cdot q) p' \cdot q
 \nn \\ &&
-(D-3) p'^2 p \cdot q+(D-2) p'^2 q^2-(D-3) p \cdot p'
q^2)\big]U_{A}(p,q,p') \nn
\\ && + \big[ (m^2-q^2) \left(p'^2
q^2-(p' \cdot q)^2\right)\big]U_{B}(p,q,p') \nn \\ && +
\big[(p'^2-m^2) \left((p' \cdot q)^2-p'^2
q^2\right)\big]U_{C}(p,q,p') \nn \\ && + \big[(D-4) (p' \cdot
q)^2-(D-3) (p \cdot p'+p \cdot q) p' \cdot q+(D-3) p'^2 p \cdot q
 \nn \\ &&
-(D-4) p'^2 q^2 +(D-3) p
\cdot p' q^2 \big]\widetilde{I}_{0}(q -p,p' - p)                     \nn \\
&& - (D - 3)\big[(p' \cdot q)^2-(p \cdot p'+p \cdot q) p' \cdot q
+p'^2 (p \cdot q-q^2)+p \cdot p' q^2 \big]\times \nn \\  &&
\big[\widetilde{I}_{A}(q -p,p' - p)  +\widetilde{I}_{B}(q -p,p' -
p)\big] +
(D - 3)\big[(p' \cdot q)( p \cdot p')-p'^2 p \cdot q \big]\times \nn \\
&& I_{A}(p,p') + (D -3)\big[(p' \cdot q)(p \cdot q) - p\cdot p'
q^2\big]I_{A}(p,q) \big\} \;, \eea
\bea U_{H}(p,q,p')&=&\frac{1}{2 (D-3) \left[p^2 (p' \cdot q)^2-2 (p
\cdot p')( p \cdot q) (p' \cdot q)+(p \cdot p')^2 q^2+p'^2 \left((p
\cdot q)^2-p^2 q^2\right)\right]}\nn \\ && \big \{-\big[(m^2-p^2)
((p' \cdot q) p^2-(p \cdot p')( p \cdot q))\big]U_{A}(p,q,p')\nn \\
&& + \big[(D-3) p \cdot q (p'^2 p \cdot q-p' \cdot q p^2)+(-p^2
((D-3) p'^2+p' \cdot q+3 p \cdot p' \nn \\ && -D (p' \cdot q+p \cdot
p'))-(D-1) p \cdot p' p \cdot q) q^2+m^2 (p' \cdot q (-D p^2+p^2 \nn
\\ && +(D-3) p \cdot q)-p \cdot q (-D
p \cdot p'+p \cdot p'+(D-3) p \cdot q)+(D-3) (p^2-p \cdot p')
q^2)\big]\nn \\ && \frac{U_{B}(p,q,p')}{2} + \big[m^2 (p' \cdot q
(-D p^2+p^2+(D-3) p \cdot p')+(D-3) p'^2 (p^2-p \cdot q)\nn \\ && -p
\cdot p' ((D-3) p \cdot p'-D p \cdot q+p \cdot q))+p'^2 ((D-1) p'
\cdot q p^2 \nn \\ && +(D-3) (p \cdot q-q^2) p^2-(D-1) p \cdot p' p
\cdot q)+(D-3) p \cdot p' (p \cdot p' q^2-p' \cdot q p^2) \big]\nn
\\ && \frac{U_{C}(p,q,p')}{2} + \big[(p \cdot p')(p\cdot q)-
p'\cdot q p^2 \big]\widetilde{I}_{0}(q -p,p' - p) \nn \\ && +
\frac{D - 3}{2}\big[ p \cdot q^2-p' \cdot q p \cdot q-p \cdot p' p
\cdot q+p' \cdot q p^2-p^2 q^2+p \cdot p' q^2\big] \nn \\ &&
\widetilde{I}_{A}(q -p,p' - p)  +\frac{D - 3}{2} \big[p' \cdot q
(p^2-p \cdot p')+p \cdot
p' (p \cdot p'-p \cdot q)+ \nn \\
&& p'^2 (p \cdot q-p^2) \big]\widetilde{I}_{B}(q -p,p' - p) +
\frac{D - 3}{2} \big[p'^2p^2
-(p\cdot p')^2 \big]I_{B}(p,p')  \nn \\
&& + \frac{D - 3}{2}\big[p^2q^2-(p \cdot q)^2 \big]I_{B}(p,q) +
\frac{D -
3}{2}\big[p' \cdot q p\cdot q-p\cdot p'q^2 \big]I_{A}(p',q) \nn \\
&& + \frac{D - 3}{2}\big[p' \cdot q p\cdot p'-p'^2 p\cdot
q\big]I_{B}(p',q) \big \}  \;. \\ \nn \\  
U_{G}(q,p,p')&=&
U_{I}(p',q,p)=U_{D}(p,q,p') \;, \quad U_{E}(p,q,p')=U_{H}(p',q,p)
\quad U_{F}(p,q,p')=U_{H}(q,p,p') \;.
\eea  \\

\noindent
{\bf  The $V^{\mu \nu}(p,q,p')$ Integral~:} \\

\noindent Similarly to the previous case, the  $V^{\mu\nu}$ integral
can be expanded in terms of $U_{0}, V_{D}, V_{E}$ ,$V_{F}$, $V_{G}$,
$V_{H}$ and $V_{I}$ as follows~: 
\bea
V_{\mu\nu}^{(2)}(p,q,p')&=&\frac{i\pi^{2}}{2}\bigg[\frac{g_{\mu\nu}}{D}
U_{0}(p,q,p')
+ \bigg(p_{\mu}p_{\nu} - g_{\mu\nu}\frac{p^{2}}{D}\bigg)V_{D} +
\bigg(p_{\mu}q_{\nu} + q_{\mu}p_{\nu} -
g_{\mu\nu}\frac{2(p \cdot q)}{D}\bigg)V_{E} \nn\\
&& \hspace{1cm} + \bigg(p_{\mu}p'_{\nu} + p'_{\mu}p_{\nu} -
g_{\mu\nu}\frac{2(p \cdot p')}{D}\bigg)V_{F} +\bigg(q_{\mu}q_{\nu} -
g_{\mu\nu}\frac{q^{2}}{D}\bigg)V_{G} \nn \\ && \hspace{1cm} +
\bigg(q_{\mu}p'_{\nu} + p'_{\mu}q_{\nu} - g_{\mu\nu}\frac{2(p' \cdot
q)}{D}\bigg)V_{H} + +\bigg(p'_{\mu}p'_{\nu} -
g_{\mu\nu}\frac{p'^{2}}{D}\bigg)V_{I} \bigg]\;,
\eea
where  the independent scalar integrals are~:
\bea V_{D}(p,q,p')&=&\frac{1}{2 (D-3) \left[p^2 (p'
\cdot q)^2-2 (p \cdot p')( p \cdot q )(p' \cdot q)+(p \cdot p')^2
q^2+p'^2 \left((p \cdot q)^2-p^2 q^2\right)\right]}\nn
\\ && \big \{\big[ (D-2) p^2 (p' \cdot q)^2-(D-3) (p'^2 p \cdot q+p
\cdot p' q^2) p' \cdot q+p'^2 ((D-3) (p \cdot p'+p \cdot q) \nn \\
&& -(D-2) p^2) q^2+m^2 \big(-(D-2) (p' \cdot q)^2+(D-3) (p \cdot
p'+p \cdot q) p' \cdot q \nn \\ && -(D-3) p'^2 p \cdot q+(D-2) p'^2
q^2-(D-3) p \cdot p' q^2\big)\big]V_{A}(p,q,p') \nn \\ && +
\big[(m^2-q^2) \left(p'^2 q^2-(p' \cdot q)^2\right)
\big]V_{B}(p,q,p')+ \big[(p'^2-m^2) \left((p' \cdot q)^2-p'^2
q^2\right)\big] \nn \\ && V_{C}(p,q,p') + \big[(D-2) (p' \cdot
q)^2-(D-3) (p \cdot p'+p \cdot q) p' \cdot q+(D-3) p'^2 p \cdot q
\nn \\ && -(D-2) p'^2 q^2+(D-3) p \cdot p' q^2 \big]U_{A}(p,q,p') +
\big[(p' \cdot q)^2-p'^2 q^2 \big][U_{B}(p,q,p') \nn \\ && +
U_{C}(p,q,p')] + (D - 3)\big[p' \cdot q p \cdot p'-p'^2 p \cdot q
\big]J_{A}(p,p') \nn \\ && + (D - 3)\big[p' \cdot q p \cdot q-
p\cdot p' q^2 \big]J_{A}(p,q) + 2\big[p'^2 q^2-(p' \cdot q)^2
\big]U_{0}(p,q,p') \big \}  \;, \\ \nn \\ \nn
V_{H}(p,q,p')&=&\frac{1}{2 (D-3)
\left[p^2 (p' \cdot q)^2-2 (p \cdot p')( p \cdot q )(p' \cdot q)+(p
\cdot p')^2 q^2+p'^2 \left((p \cdot q)^2-p^2 q^2\right)\right]}\nn
\\ && \big \{-\big[(m^2-p^2) (p' \cdot q
p^2-p \cdot p' p \cdot q)\big]V_{A}(p,q,p') \nn \\ && + \big[(D-3) p
\cdot q (p'^2 p \cdot q-p' \cdot q p^2)+(-p^2 ((D-3) p'^2+p' \cdot
q+3 p \cdot p' \nn \\ && -D (p' \cdot q+p \cdot p'))-(D-1) p \cdot
p' p \cdot q) q^2+m^2 (p' \cdot q (-D p^2+p^2+(D-3) p \cdot q)- \nn
\\ && p \cdot q (-D
p \cdot p'+p \cdot p'+(D-3) p \cdot q)+(D-3) (p^2-p \cdot p') q^2)
\big]\frac{V_{B}(p,q,p')}{2} \nn \\ && \big[m^2 (p' \cdot q (-D
p^2+p^2+(D-3) p \cdot p')+(D-3) p'^2 (p^2-p \cdot q) -p \cdot p'
((D-3) p \cdot p' \nn \\ &&-D p \cdot q+p \cdot q))+p'^2 ((D-1) p'
\cdot q p^2+(D-3) (p \cdot q-q^2) p^2-(D-1) p \cdot p' p \cdot q)
\nn \\ && +(D-3) p \cdot p' (p \cdot p' q^2-p' \cdot q p^2)
\big]\frac{V_{C}(p,q,p')}{2} + \big[p' \cdot q p^2-p \cdot p' p
\cdot q\big]U_{A}(p,q,p') \nn \\ && \big[(D-1) p' \cdot q p^2-(D-3)
p' \cdot q p \cdot q+p \cdot q (-D p \cdot p'+p \cdot p' +(D-3) p
\cdot q) \nn \\ && +(D-3) (p \cdot p'-p^2) q^2
\big]\frac{U_{B}(p,q,p')}{2} + \big[(D-1) p' \cdot q p^2-(D-3) p'
\cdot q p \cdot p' \nn \\ && +(D-3) p'^2 (p \cdot q-p^2)+p \cdot p'
((D-3) p \cdot p'-D p \cdot q+p \cdot q)
\big]\frac{U_{C}(p,q,p')}{2} \nn \\ && + \frac{(D - 3)}{2}\big[p'^2
p^2-(p \cdot p')^2 \big]J_{B}(p,p') + \frac{(D - 3)}{2}\big[p^2
q^2-(p \cdot q)^2 \big]J_{B}(p,q) \nn \\ && + \frac{(D - 3)}{2}\big[
p' \cdot q p \cdot q-p \cdot p' q^2 \big]J_{A}(p',q) + \frac{(D -
3)}{2}\big[p' \cdot q p \cdot p'-p'^2 p\cdot q \big]J_{B}(p',q) \nn \\
&& + 2\big[p \cdot p' p \cdot q-p' \cdot q p^2 \big]U_{0}(p,q,p')
\big \}  \;,  \\ \nn \\ V_{G}(q,p,p') &=&V_{I}(p',q,p)=
V_{D}(p,q,p') \quad V_{E}(p,q,p')=V_{H}(p',q,p) \quad
V_{F}(p,q,p')=V_{H}(q,p,p') \;. \eea
%
%
%


\end{document}